# The Role of Edge Resonant Magnetic Perturbations in Edge-Localized-Mode Suppression and Density Pump-out in low-collisionality DIII-D Plasmas

Q.M. Hu[1], R. Nazikian[1], B.A. Grierson[1], N.C. Logan[1], and C. Paz-Soldan[2], and Q. Yu[3]

[1]*Princeton Plasma Physics Laboratory, Princeton NJ 08543-0451, USA*
[2] *General Atomics, PO Box 85608, San Diego, California 92186-5608, USA*
[3] *Max-Plank-Institut fur Plasmaphysik, 85748 Garching, Germany*

Email: qhu@pppl.gov

**Abstract**: Two-fluid nonlinear MHD simulations using the TM1 code demonstrate that the formation of magnetic islands at the top and bottom of the H-mode pedestal, together with the strong screening of resonant fields in the gradient region of the pedestal, can account for ELM suppression and density pump-out by $n$ = 2 Resonant Magnetic Perturbations (RMPs) in low-collisionality DIII-D ITER Similar Shape (ISS) plasmas. Using experimentally relevant transport coefficients, neoclassical resistivity, electron collisionality, and RMP amplitudes, nonlinear MHD simulations reproduce the observed level of density reduction (density pump-out) in DIII-D due the formation of narrow magnetic islands and resulting enhanced collisional transport in the resistive foot of pedestal. For large amplitude RMPs ($B_r/B_t>1\times10^{-4}$) simulations predict field penetration and pressure reduction at the top of the pedestal consistent with experimental observations at the onset of ELM suppression. The predicted reduction in the height and width of the pedestal by magnetic island enhanced transport provides a quantitative mechanism for the stabilization of the Peeling-Ballooning Mode (PBM). Importantly, these simulations predict strong screening of resonant fields in the steep gradient region of the pedestal due to strong **E**×**B** and diamagnetic flows. However, if the plasma resistivity is made artificially larger (~10X) than neoclassical, the simulations predict magnetic stochasticity throughout the plasma edge and the collapse of the pedestal due to the reduction in the penetration threshold with increasing resistivity. A scaling relation for resonant field penetration at the pedestal top, using several hundred nonlinear simulations, reproduces the density and **E**×**B** dependence of the ELM suppression threshold observed in DIII-D. Simulations using ITER model equilibria suggest that the penetration threshold at the top of the ITER pedestal will be much lower than in DIII-D due to the anticipated lower perpendicular flow velocities in ITER, resulting in the weaker screening of resonant fields.

## 1. Introduction

For a tokamak fusion reactor, the control of edge-localized-modes (ELMs) in high-confinement mode (H-mode) plasmas is an essential requirement to avoid excessive heat and particle fluxes onto the plasma facing components (PFCs). ITER, the fusion reactor being constructed in southern France, must operate with strongly mitigated or suppressed ELMs at high power in order to avoid excessive erosion and impurity generation [1]. One method to completely suppress ELMs is with the use of edge resonant magnetic perturbations (RMPs) [1] that was first demonstrated on the DIII-D tokamak [2] and thereafter demonstrated on several tokamaks worldwide. The RMPs also produce enhanced particle transport without a significant impact on the pedestal temperature [3], called density pump-out, and understanding the transport effect of RMPs is essential to predict ITER performance. RMPs have been used to mitigate [4–6] and suppress [3,7–9] ELMs in many tokamaks, however, the causes of ELM suppression and pump-out are still not well understood. Understanding the physics mechanisms responsible for ELM suppression and density pump-out by RMPs has been a longstanding issue for RMP ELM control with important implications for the design of ELM control coils in ITER [10–34]. Various mechanisms have been proposed to explain the phenomenology of RMP effects on the H-mode pedestal. These include the role of magnetic stochasticity in pedestal transport [2,10–16], coupling to ballooning modes in ELM mitigation and suppression [17–19], turbulence effects on edge transport [15,20–22], magnetic island formation and resonant $q_{95}$ windows of ELM suppression [3,8,15,23], and non-ambipolar transport for the modification of the edge electric field, to name a few [32–34].

It is most likely that all these mechanisms have some regime of validity in the description of RMP ELM suppression and density pump-out, however a key question is whether a predictive model of ELM suppression and pump-out can be constructed that (a) reproduces much of the phenomenology and parametric trends in the existing data, and (b) provides a physics basis to predict ELM suppression and pump-out in low-collisionality ITER high fusion power plasmas. To date no one model has demonstrated agreement with the key trends observed in experiment including the ubiquitous low density threshold [29,30] and high toroidal rotation [30] required to access ELM suppression, the relatively weak effect of the RMP on the temperature relative to the density during pump-out [3], and the preservation of the edge-transport-barrier (ETB) during density pump-out and ELM suppression.

In a recent paper we showed that two-fluid nonlinear MHD simulations using DIII-D relevant profiles, resistivity, collisionality, transport coefficients and RMP amplitudes, reproduce the key trends observed in DIII-D RMP experiments including ELM suppression and density pump-out [35]. The analysis shows that strong density pump-out with weak temperature reduction occurs at low collisionality from resonant field penetration at the pedestal foot, whereas ELM suppression results from resonant field penetration and enhanced transport at the pedestal top. The simulations also predict the strong screening of resonant fields in the gradient region of the pedestal, consistent with the experimental observation of strong temperature gradients during ELM suppression and density pump-out in DIII-D. In this paper we extend our results to include the followings: (1) MHD simulations of plasmas with varying density, demonstrating the density dependence of pump-out due to collisionality at the foot of the pedestal; (2) detailed comparison of MHD simulation with the observed **E**×**B** profile, showing areas of agreement and disagreement between experiment and simulation; (3) a scaling relation for resonant penetration at the pedestal top based on MHD simulation and comparisons with experiment, revealing consistency with the density and **E**×**B** dependence of the ELM suppression threshold in DIII-D; (4) nonlinear MHD predictions of $n$ = 3 resonant field penetration in ITER, indicating that the threshold for field penetration at the top of the ITER pedestal could be significantly lower than in DIII-D; (5) prediction of a hysteresis effect for resonant field penetration in the pedestal, consistent with experimental observations, and (6) a sensitivity study for the dependence of the penetration threshold on transport coefficients used in the simulations.

Here we present nonlinear two-fluid MHD simulations over a wide range of experimental parameters to demonstrate that the key phenomenology and trends in ELM suppression and density pump-out can be attributed to magnetic island formation and their effects on transport. We employ the cylindrical TM1 code for these simulations due to its computational optimization for low collisionality plasma conditions relevant to modern fusion experiments. The initial conditions for TM1 are profiles and transport coefficients taken before the RMP is applied, together with helical magnetic boundary conditions obtained from perturbative 3D equilibrium calculations in full plasma geometry. The TM1 model then computes the bifurcation threshold for resonant field penetration and the modified profiles resulting from magnetic island formation due to enhanced parallel neoclassical transport.

Our choice of nonlinear MHD model is guided by the need to efficiently compute the plasma response to RMPs over a wide range of experimentally relevant parameters in order to infer trends that can be compared directly to experiment. Our study is therefore complementary to computationally much more intensive and sophisticated gyro-kinetic [21,36,37] and extended MHD [11,17–19] approaches that can only address limited cases and therefore have difficulty inferring trends that can be compared to experiment. The successes and limitations of our approach are discussed.

The paper is organized as follows. The basic phenomenology of ELM suppression and pump-out by $n = 2$ RMPs is presented in section 2. In section 3.1, the nonlinear two-fluid TM1 model is introduced together with a detailed discussion of the model inputs and boundary conditions. Section 3.2 presents results from TM1 simulations of resonant field penetration at the top of the pedestal together with measurements at the onset of ELM suppression. Section 3.3 presents simulations of density pump-out due to field penetration at the foot of the pedestal and comparisons with experiment. Section 3.4 introduces the scaling of resonant field penetration at the top of the pedestal from MHD simulations and its consistency with the density and **E**×**B** rotation dependence of ELM suppression in DIII-D experiments. Sections 3.5 shows how hysteresis in the ELM suppression threshold is consistent with MHD simulation for the penetration threshold and screening of resonant fields at the pedestal top. Section 3.6 shows the sensitivity of the penetration threshold on transport coefficients. The limitation of the TM1 model is discussed and a summary of results presented in section 4.

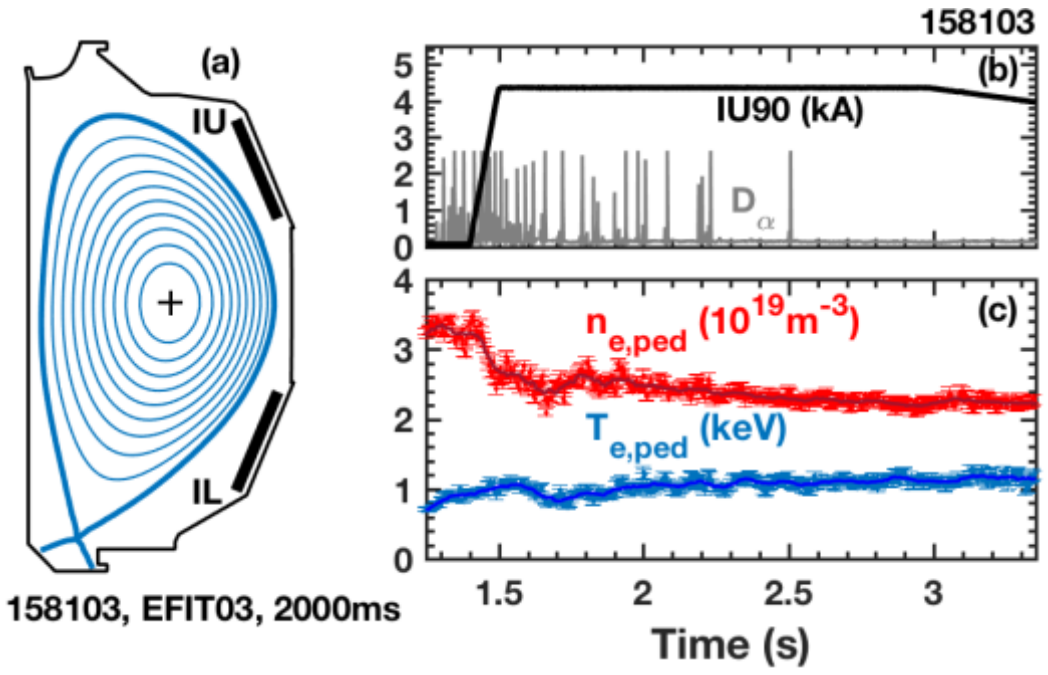

**Figure 1**. Typical ELM suppression by $n = 2$ RMP in a low collisionality ITER Similar Shape (ISS) plasma in DIII-D: (a) ITER shaped equilibrium cross-section and the location of upper I-coils (IU) and lower I-coils (IL). Time traces of (b) $D_\alpha$ light and RMP coil current, and (c) pedestal density $n_{e,\,\mathrm{ped}}$ (red) and electron temperature $T_{e,\,\mathrm{ped}}$ (blue).

## 2. ELM suppression by $n = 2$ RMPs

The low-collisionality DIII-D discharges presented here are configured in the ITER-similar-shape (ISS), with plasma current $I_p$ = 1.37 MA, toroidal field $B_T$ = 1.94 T, minor radius $a$ = 0.6 m, internal inductance $l_i \approx 1$, neutral beam injected power ≈ 6 MW, normalized beta $\beta_N \approx 2.5$, central electron cyclotron resonance heating ≈1 MW, normalized electron pedestal collisionality $\nu^*_e \approx$ 0.1-0.5 and edge safety factor $q_{95} \approx 4.1$ [23]. Figure 1 shows the ITER shaped equilibrium together with time traces for the edge Balmer alpha ($D_\alpha$) line emission and internal coil (I-coil) current = 4.5 kA [38]. The I-coils are configured for toroidal mode number $n = 2$ [39]. A long period of ELM suppression is sustained, characterized by the absence of transient spikes in the $D_\alpha$ signal. Also shown is the evolution of the plasma density and temperature at the top of the pedestal in figure 1(c). The RMP induced density pump-out can be seen in the interval 1.4 s < $t$ < 1.8 s before ELM suppression with minimal effect on the electron temperature. This is a typical case showing density pump-out before ELM suppression in DIII-D.

For the cases studied in this paper, the resonant field strength was varied slowly in order to identify the nonlinear dynamics specific to ELM suppression and pump-out [15,23]. The resonant field strength is sinusoidally varied by controlling the relative phase $\Delta\Phi_{UL}$ of the $n = 2$ current between the two rows of I-coils. The $n = 2$ field in the upper row of coils is rotated toroidally at 1 Hz while the $n = 2$ field in the lower row is held fixed with an I-coil current of 4 kA, producing a sinusoidal resonant ($m/n$ = 8/2) magnetic field perturbation on the plasma surface $\psi_N = 1$ shown in figure 2(a). Here, $B_r$ is the radial magnetic perturbation at the plasma surface ($\psi_N = 1$) in the experiment. The magnetic probe suite on DIII-D also allow direct measurement of $n = 2$ magnetic responses at the plasma midplane on the low-field side (LFS) and high-field side (HFS) of the plasma [40].

The magnitude of the $m/n$ = 8/2 resonant field strength on the plasma boundary $\psi_N$ = 1 is calculated using the full geometry toroidal ideal-MHD Generalized Perturbed Equilibrium Code (GPEC) [41] as shown in figure 2(a). GPEC calculates perturbatively the total (vacuum + plasma response) field produced by the I-coil currents. There are many such helical fields in the edge for different mode numbers ($m/n$ = 8/2, 9/2, 10/2, 11/2, etc), however, they all vary similarly so for simplicity we only show the $m/n$ = 8/2 component. For these shots, GPEC predicts the maximum edge resonant field at $\psi_N = 1$ at $\Delta\Phi_{UL} \approx 0°$, and the minimum near $\Delta\Phi_{UL} \approx 180°$. The locations of $m/n$ = 7/2, 8/2, 9/2, 10/2 and 11/2 rational surfaces are $\psi_N$ = 0.888, 0.928, 0.958, 0.978 and 0.992, based on kinetic equilibrium (EFIT) analysis. The top of the pedestal location is obtained using Tanh fits to the electron temperature and is at $\psi_N$ = 0.935. Thus, we use $q_{95}$ as a convenient proxy for the safety factor near the top of the pedestal.

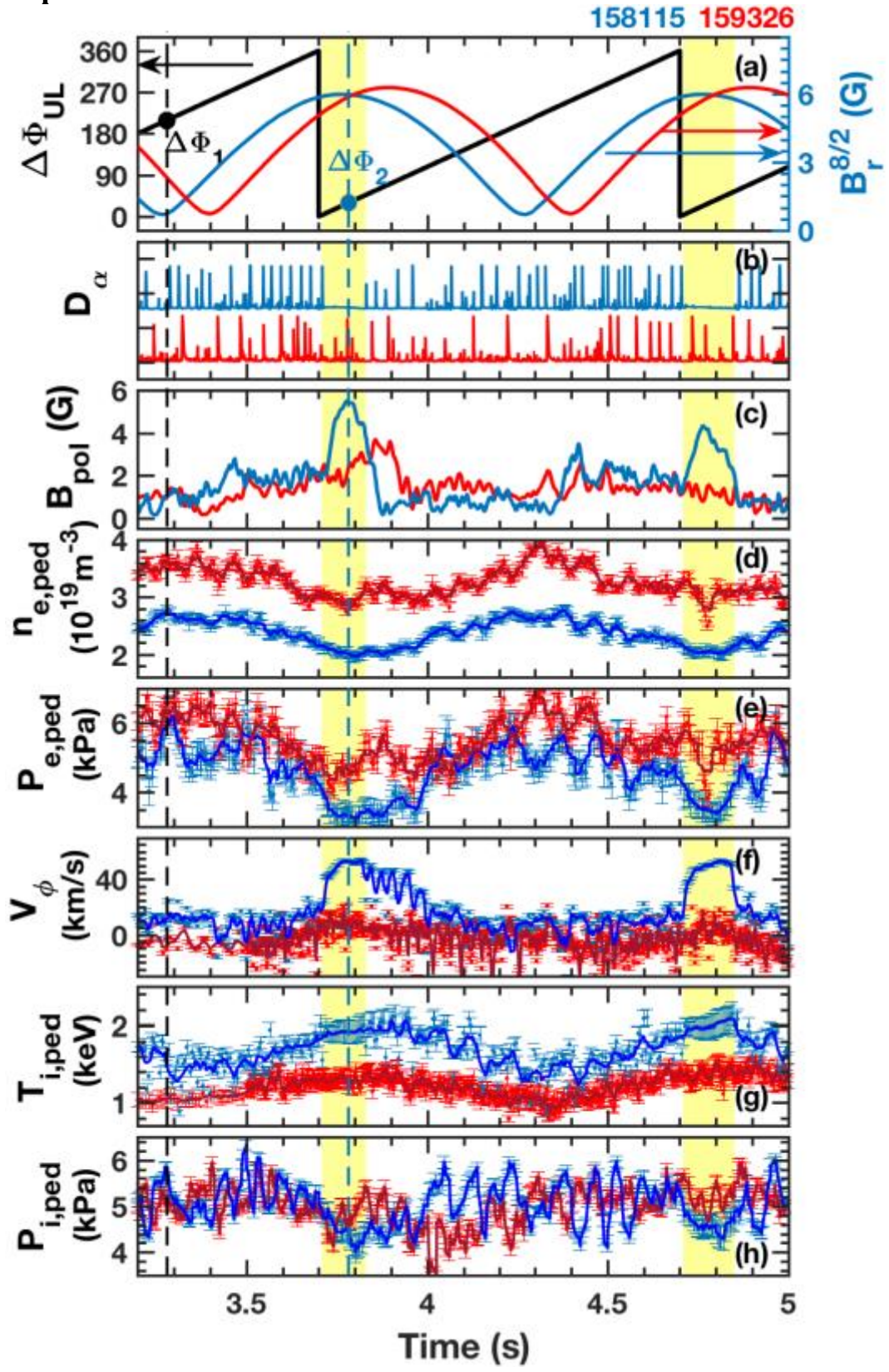

**Figure 2**. For two similar ELM control discharges, with (shot #158115, blue curves) and without (shot #159326, red curves) ELM suppression during application of slowly varying RMPs: (a) Upper and lower row I-coil relative phase $\Delta\Phi_{UL}$, and RMP strength for $m/n$ = 8/2 component calculated by GPEC (including the total of vacuum and ideal plasma response from the I coils). (b) $D_\alpha$ light near the outer strike point. (c) $n$ = 2 plasma poloidal magnetic field response amplitude $|B_{pol}|$ on the high field side midplane magnetic sensors. Pedestal (d) density $n_{e,\,ped}$ and (e) electron pressure $P_{e,ped}$. Edge impurity velocity (f) in the co-$I_p$ direction, pedestal (g) ion temperature $T_{i,\,ped}$ and (h) ion pressure $P_{i,ped}$.

In this paper we focus on recovering the key trends in the experimental data, as shown in figure 2. The first key trend is the abruptness of the pedestal response to the RMP at the onset of ELM suppression [15]. The second key trend is the almost universal observation of a low-density threshold required for ELM suppression in low-collisionality DIII-D plasmas [30]. The third trend is the uniform variation of the plasma density (called density pump-out) with a relatively weak effect on the pedestal temperature in response to the uniform variation of the RMP strength [23].

A typical ELM control experiment using $n$ = 2 RMPs is shown in figure 2, which illustrates many of these features generic to RMP effects in the pedestal. The first discharge in figure 2 (#158115) has a low pedestal density and transitions to ELM suppression in narrow time windows indicated by the yellow shaded bands. The second discharge (#159326) is at a higher pedestal density and does not exhibit ELM suppression. Yet both discharges exhibit uniformly varying density (pump-out).

Near $\Delta\Phi_{UL} \approx 0°$ (referred to as the even parity I-coil configuration) the plasma transitions to ELM suppression in narrow intervals during the peak in the edge resonant response for the low-density discharge #158115. The transition to ELM suppression is characterized by the absence of $D_\alpha$ spikes (figure 2(b)), the abrupt increase in the poloidal magnetic response on the inner wall (figure 2(c)), and an increase in the edge carbon impurity toroidal rotation (figure 2(f)). The pedestal pressure also drops suddenly in the intervals of ELM suppression as shown in figure 2(e). Here, the pedestal density $n_{e,ped}$ and pressure $P_{e,ped}$ are obtained using hyperbolic tangent fits of edge Thomson scattering profiles [42]. The carbon-VI toroidal velocity $V_\phi$ near the normalized poloidal flux $\psi_N \approx 0.94$ is measured using charge exchange emission [43].

We note from figure 2 that the ion temperature exhibits no reduction (even increases slightly) in the intervals of ELM suppression. While it would be beneficial to have main ion measurements in the pedestal, these were not available for our study. We note that the higher density plasma does not transition to ELM suppression [44] and that the experimentally determined density threshold for ELM suppression ($n_{e,ped} < 3\times10^{19}$m$^{-3}$) is qualitatively similar to the density threshold for core resonant field penetration [45] in DIII-D plasmas.

In contrast to the sudden changes at the onset of ELM suppression, density pump-out is more gradual. The density pump-out, indicated by the slow variation of the pedestal density $n_{e,\mathrm{ped}}$ in figure 2(d), varies opposite to the strength of the applied resonant field [23]. Both discharges exhibit similar pump-out behavior although only the low-density discharge exhibits ELM suppression.

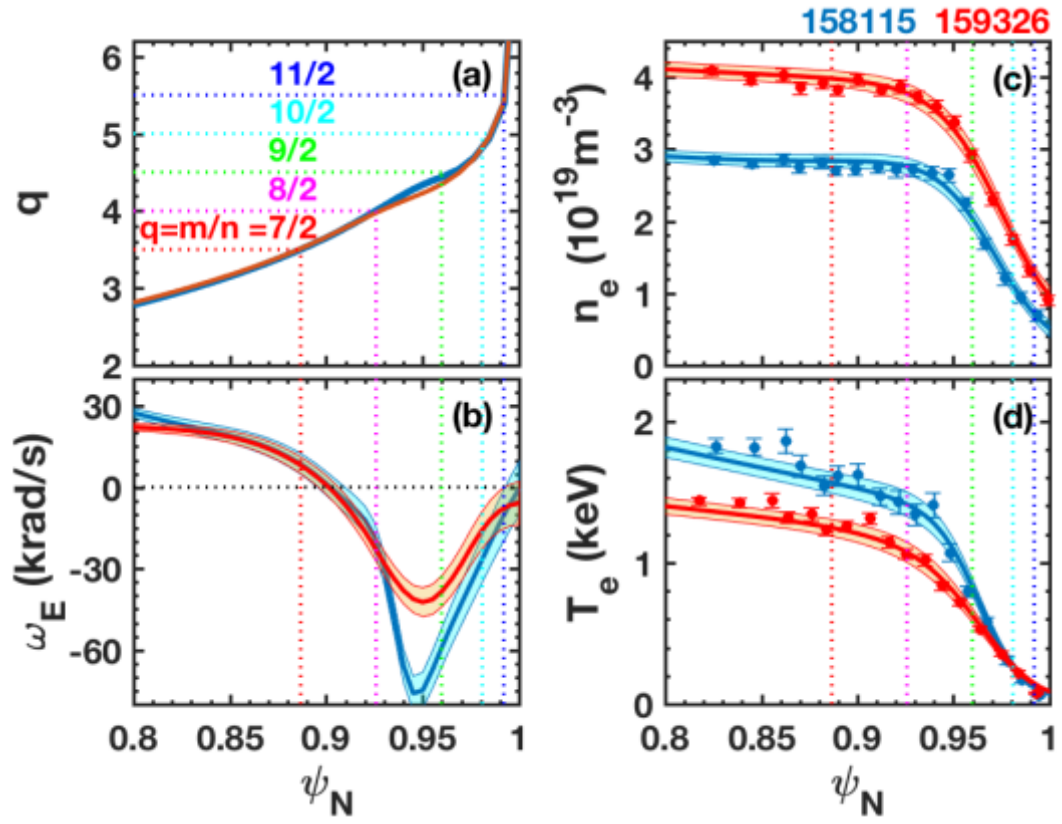

**Figure 3**. Equilibrium profiles of (a) safety factor $q$, (b) **E**×**B** rotation $\omega_E$, (c) electron density $n_e$ and (d) temperature $T_e$ overlaid with Thomson scattering data (symbols), taken at the time of minimum RMP amplitude for shot 158115 (blue, 3.25 s) and 159326 (red, 3.4 s). The rational surfaces $q = m/n = 7/2$, 8/2, 9/2, 10/2 and 11/2 are located at $\psi_N = 0.888$, 0.928, 0.958, 0.978 and 0.992 as shown by the dotted lines, respectively. The $q$-profile is obtained from kinetic equilibrium analysis using the measured profiles. The uncertainty of the profiles is also shown by the overlay of raw data, error bars and the band of uncertainty in the fitted profiles.

The plasma profiles at the minimum RMP amplitude (at $t = 3.25$ s and 3.4 s in figure 2) are shown in figure 3 for the two discharges with their uncertainties. For the kinetic equilibrium, the convergence error of the Grad-Shafranov equation is about $8\times10^{-9}$, and the total $\chi^2$ ($\chi$-squared goodness-of-fit parameter against the experimental measurements) is about 30, including MSE, magnetic measurements and pressure. Here, we point out that the level of the uncertainty in these profiles are similar for different plasmas studied in our paper. We denote these as the "initial" profiles when the RMP amplitude is negligible and these are the input profiles used for our nonlinear simulations. The **E**×**B** profiles shown in figure 3 ($\omega_E = E_r/|RB_\theta|$) are obtained from charge exchange measurements of the carbon-VI impurity.

## 3. TM1 and GPEC simulation

### 3.1. Theoretical model

For our analysis we use the TM1 code [46–48], which has recently been improved to perform high-resolution simulations with multiple helical boundary conditions relevant to the DIII-D pedestal [See Appendix A for a more detailed description of the TM1 model]. TM1 is a nonlinear time-dependent two-fluid MHD code with cylindrical geometry and circular cross-section. TM1 includes the nonlinear coupling of harmonics of each helicity (i.e. for the $m/n = 8/2$ it computes the coupling to $m/n = 16/4$, 24/6, 32/8, etc).

The TM1 code solves the motion equation which evaluates the torque balance between the electromagnetic (**J**×**B**) torque due to the RMP and the plasma viscosity in the motion equation (A3) in Appendix A. The torque balance governs the bifurcation from screening to penetration of resonant fields in the pedestal, which is also sensitive to diamagnetic drifts through Ohm's Law in equation (A1). The electron density and temperature are self-consistently evaluated through the continuity equation (A4) and the energy transport equation (A5). Physically, resonant field penetration enhances parallel neoclassical transport across the magnetic island, leading to enhanced particle and thermal transport at rational surfaces. The enhanced neoclassical particle and thermal transport are solved self-consistently with the penetration or screening of resonant fields in the TM1 model. Due to two-fluid effect, particle and energy transport are coupled to the motion equation that determines field penetration or screening. As a result, the penetration threshold depends on plasma parameters such as **E**×**B** rotation, electron density and temperature, diamagnetic drifts, RMP strength and anomalous transport coefficients (plasma viscosity, particle and heat diffusivity) determined by transport analysis of experimental profiles. Previously, the TM1 code has reproduced the conditions for error field penetration in the core of TEXTOR [49,50] and DIII-D [51]. The TM1 code has also studied magnetic island induced heat and particle transport in the ASDEX-upgrade [46,47,52,53], DIII-D [51] and J-TEXT [54,55] tokamaks.

The cylindrical geometry utilized in the TM1 model is substantially different to the toroidal strongly shaped magnetic geometry in the DIII-D experiment. To minimize the associated effect due to this difference, the fully toroidal MHD code, the Generalized Perturbed Equilibrium Code (GPEC) [41] is used to calculate the perturbed 3D magnetic equilibrium that forms the boundary condition for TM1 simulations. Here, GPEC is an upgrade of the IPEC code [56] to include nonideal effects, however only the ideal version of GPEC is used (equivalent to IPEC). GPEC calculates perturbatively the total (vacuum plus ideal MHD kink) 3D magnetic response of the plasma to the I-coils in full toroidal strongly shaped magnetic geometry for a given toroidal mode number, in this case $n$ = 2. The calculated harmonics of the $n$ = 2 RMP that are resonant to rational surfaces in the edge region of the plasma are then used as the boundary condition for the TM1 simulations at $\psi_N$ = 1. The GPEC model takes as input a kinetically constrained equilibrium from DIII-D profile measurements and EFIT analysis before the application of the RMP. By doing so, we take into account the effect of the actual magnetic geometry on the helical boundary conditions for TM1 simulations. The GPEC code has been well validated in DIII-D for $n$ = 2 RMPs in plasmas very similar to the ones used for TM1 simulation [23].

TM1 also requires plasma profiles, transport coefficients, resistivity and collisionality as initial conditions to solve the system of equations A1-A5. The profiles of electron density, temperature and **E**×**B** rotation vs poloidal flux come from the same kinetic equilibrium used for GPEC analysis. Additional parameters including the particle diffusivity, electron thermal diffusivity, neoclassical resistivity and electron collisionality are obtained from TRANSP [57] power and particle balance calculations before the RMP is applied (the profiles shown in figure 3 for the two discharges in figure 2). TM1 then computes the evolution of the tearing-mode eigenfunction and the plasma response ($E_r$, $n_e$, $T_e$) for either unstable or linearly stable driven magnetic islands.

Dedicated numerical methods are utilized in TM1 to reduce the numerical error associated with large values of the magnetic Reynolds number $S$ and $\chi_\parallel/\chi_\perp$ [58]. This attribute makes TM1 particularly useful for modeling the nonlinear response of DIII-D and ITER low-collisionality plasmas to RMPs using experimentally derived parameters such as the plasma resistivity**.** TM1 has been used previously for modeling drift tearing modes [48], resonant field penetration, particle and energy transport in the core of TEXTOR [46], DIII-D [51] and J-TEXT [59] tokamaks.

In this study, multiple helicity ($m/n$ = 7/2, 8/2, 9/2, 10/2 and 11/2) magnetic boundary conditions at $\psi_N$ = 1 are included in the TM1 simulation from GPEC analysis. These helical boundary conditions are chosen in order to cover all relevant rational surfaces in the plasma edge where resonant field penetration may occur ($q_{95} \approx 4.1$ in all these $n$ = 2 RMP experiments). The calculated neoclassical resistivity [60] is used in the all these simulations, which is in the range of 1-6×$10^{-7}$ Ωm at the pedestal top and 1-6×$10^{-6}$ Ωm at the foot of the pedestal for all the plasmas studied here. The main input and output parameters for TM1 simulation are listed in Table. I.

**Table I**. Main input and output for TM1.

| | |
|---|---|
| **Input parameters** | Profiles: $q$, $\omega_E$, $n_e$, $T_e$ ← OMFIT [61] kinetic EFIT<br>Transport coefficients: $D_\perp$, $\mu$, $\chi_\perp$ ← TRANSP[52] Code<br>RMP amplitude: $B_r$ ← GPEC [41]<br>$S$, $\eta$, $\tau_R$, $\tau_A$, $d_1$, $\chi_\parallel$, $C_s$, $S_n$, $S_m$, $S_p$ ← derived from profiles |
| **Output parameters m/n=0 & ≠0** | Profiles: $j$, $\omega_E$, $n_e$, $T_e$, $\psi$<br>Derived quantities: island width ($W$), phase ($\Phi$), plasma response ($B_{pol}$, $B_r$), |

### 3.2. Pedestal-top RMP penetration and ELM suppression

The inputs for TM1 are obtained from calculated experimental parameters and kinetic equilibrium analysis at the minimum of the applied RMP. In the TM1 simulations, all the inputs are time independent except the RMP strength. The safety factor profile $q$, the initial density profile $n_e$, electron temperature profile $T_e$ and **E**×**B** rotation $\omega_E$ profile used in TM1 are shown in figure 3. The profile of the neoclassical resistivity used in TM1 is derived from the experimental profiles, as mentioned earlier. From the experimental profiles, the profile of the Reynolds number $S$ (with S = 1.1×$10^8$ at $q$ = 4 rational surface) is obtained and used in the simulation.

The TRANSP code [57] is used to obtain the effective transport coefficients from power and particle balance, including electron particle diffusivity $D_\perp$ (using the method of Porter [62]), momentum diffusion $\chi_\varphi$ (same as plasma viscosity $\mu$) and the electron heat conductivity $\chi_{\perp e}$ ($\chi_\perp$ in TM1). Here, the porter method relates the global particle to the energy confinement time in order to derive a total particle recycling flux that is used in a 1D neutral deposition calculation in TRANSP. These three transport coefficients are $D_\perp \approx \mu \approx \chi_\perp \approx$ 1 m$^2$/s at the top and bottom of pedestal region for all the plasmas studied in this paper for the profiles corresponding to the minimum of the RMP amplitude. These coefficients are taken as radially constant in the simulation. (Note that the value of $D_\perp$ at the top and foot of the pedestal is close to the magnitude obtained from previous studies in DIII-D H-mode plasmas [63]).

We point out that $D_\perp$, $\mu$, and $\chi_e$ all drop well below 1 m$^2$/s in the steep gradient region of the pedestal. However, taking these parameters as constant does not affect our results as (a) TM1 shows strong screening of resonant fields in the gradient region; and (b) TM1 maintains the initial density and temperature gradients in regions where resonant field penetration does not take place.

An example of TM1 nonlinear simulation over the full period of resonant field variation for the low-density discharge #158115 that exhibits ELM suppression is shown in figure 4. The resonant field amplitude from GPEC is shown in figure 4(a). The time evolution of the magnetic island width is shown in figure 4(b), the island toroidal phase is shown in figure 4(c), and the amplitude of $n$ = 2 poloidal magnetic perturbation at $\psi_N = 1.1$ is shown in figure 4(d). The simulated amplitude of poloidal magnetic perturbation corresponds to the location of a magnetic probe array on the HFS midplane. The RMP amplitude at $\psi_N = 1$ for each component $m/n$ = 7/2, 8/2, 9/2, 10/2 and 11/2 of the I-coil field is obtained from full geometry GPEC analysis. Only the waveform of the resonant 8/2 component is shown in figure 4(a) as the other $m/n$ components evolves similarly. The peak amplitude of all the helical field components from GPEC at $\psi_N = 1$ are: $B_r^{11/2}$ = 16 Gauss, $B_r^{10/2}$ = 11 Gauss, $B_r^{9/2}$ = 8 Gauss, $B_r^{8/2}$ = 6.2 Gauss, and $B_r^{7/2}$ = 2 Gauss.

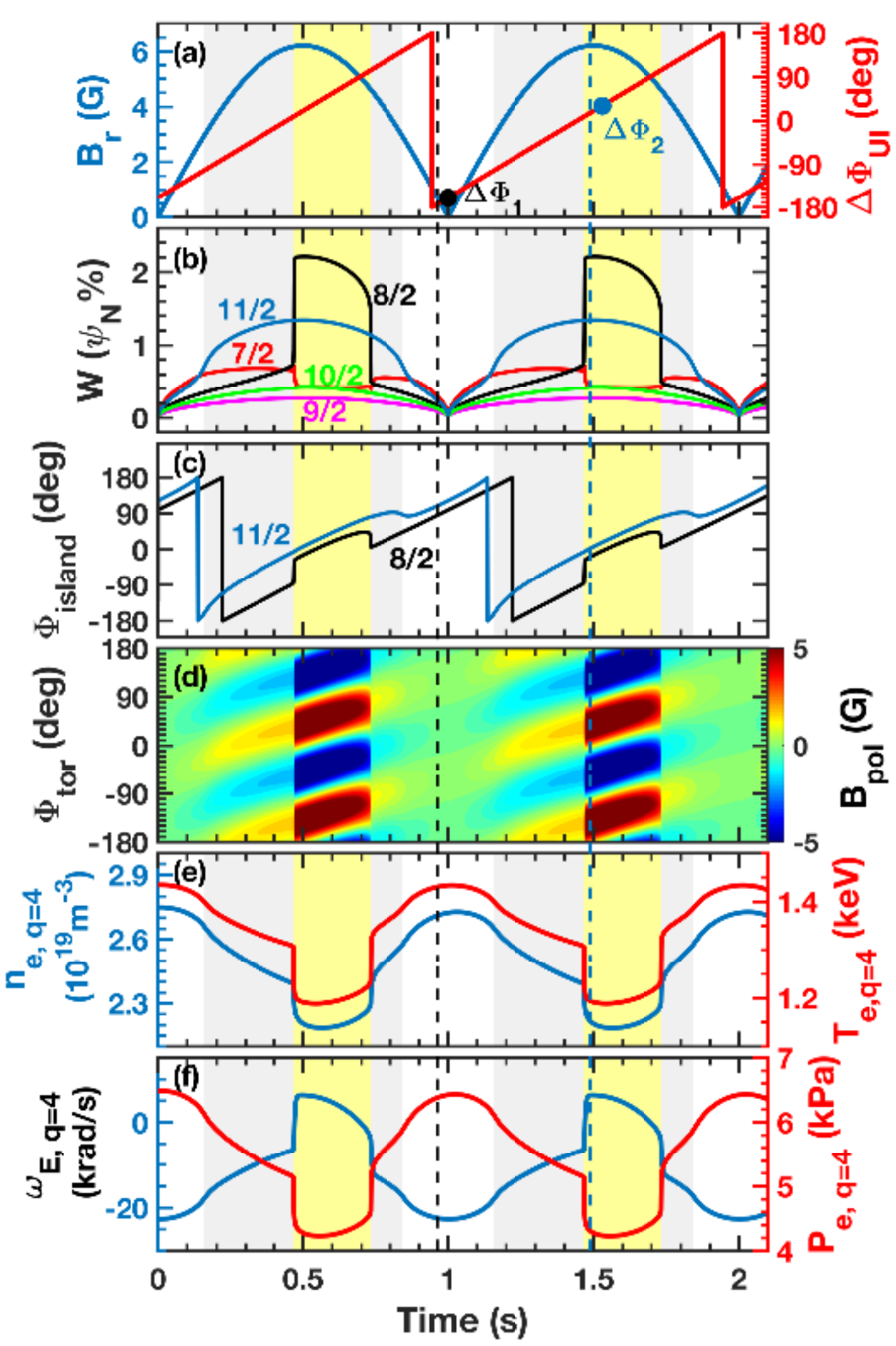

**Figure 4**. Time-domain display of the TM1 simulation for shot 158115. (a) 8/2 RMP amplitude $B_r$ from GPEC and relative phase $\Delta\Phi_{UL}$. (b) Magnetic island width W and (c) phase for $m/n$ = 8/2 and 11/2 islands. (d) Contour of $B_{pol}$ versus time and toroidal angle for $n$ = 2 at $\psi_N$ = 1.1. (e) Electron density $n_e$ and temperature $T_e$, (f) $\omega_E$ and electron pressure $P_e$ at $q$ = 4 rational surface.

A clear threshold for resonant field penetration is shown in figure 4(b) at the top and foot of the pedestal. The $m/n$ = 11/2 rational surface is located at the foot of the pedestal near the separatrix ($\psi_N \approx$ 0.99 in figure 3(a)). The width of the $m/n$ = 11/2 magnetic island at the foot of the pedestal varies smoothly with RMP amplitude (figure 4(b)). However, there is a sudden jump in the width of the 8/2 island and in the plasma response at the $q$ = 4 rational surface upon resonant field penetration at the top of the pedestal. This $m/n$ = 8/2 island is located near the top of the pedestal ($\psi_N \approx$ 0.928 in figure 3(a)).

Due to the $m/n$ = 8/2 magnetic island formation, the simulated electron density at the top of the pedestal experiences a sudden decrease (figure 4(e)). Also, the **E**×**B** rotation frequency $\omega_E$ at the $q$ = 4 rational surface experiences a sudden increase, approaching zero upon island formation. In the phase of the RMP cycle where the 8/2 RMP amplitude decreases, a second bifurcation occurs, resulting in island decay and screening. The jump in the measured magnetic perturbation on the inner wall of DIII-D reported previously [15] is consistent with the jump in the simulated poloidal field on the inner wall from TM1 (figure 4(d)) at the onset of $m/n$ = 8/2 field penetration at the $q$ = 4 surface.

In order to understand the magnetic topology due to these field penetration events, field line tracing is performed based on the TM1 solutions. The Poincaré plots of the magnetic surfaces during $m/n$ = 11/2 penetration, before and during $m/n$ = 8/2 penetration, are show in figure 5(a)-(b), respectively.

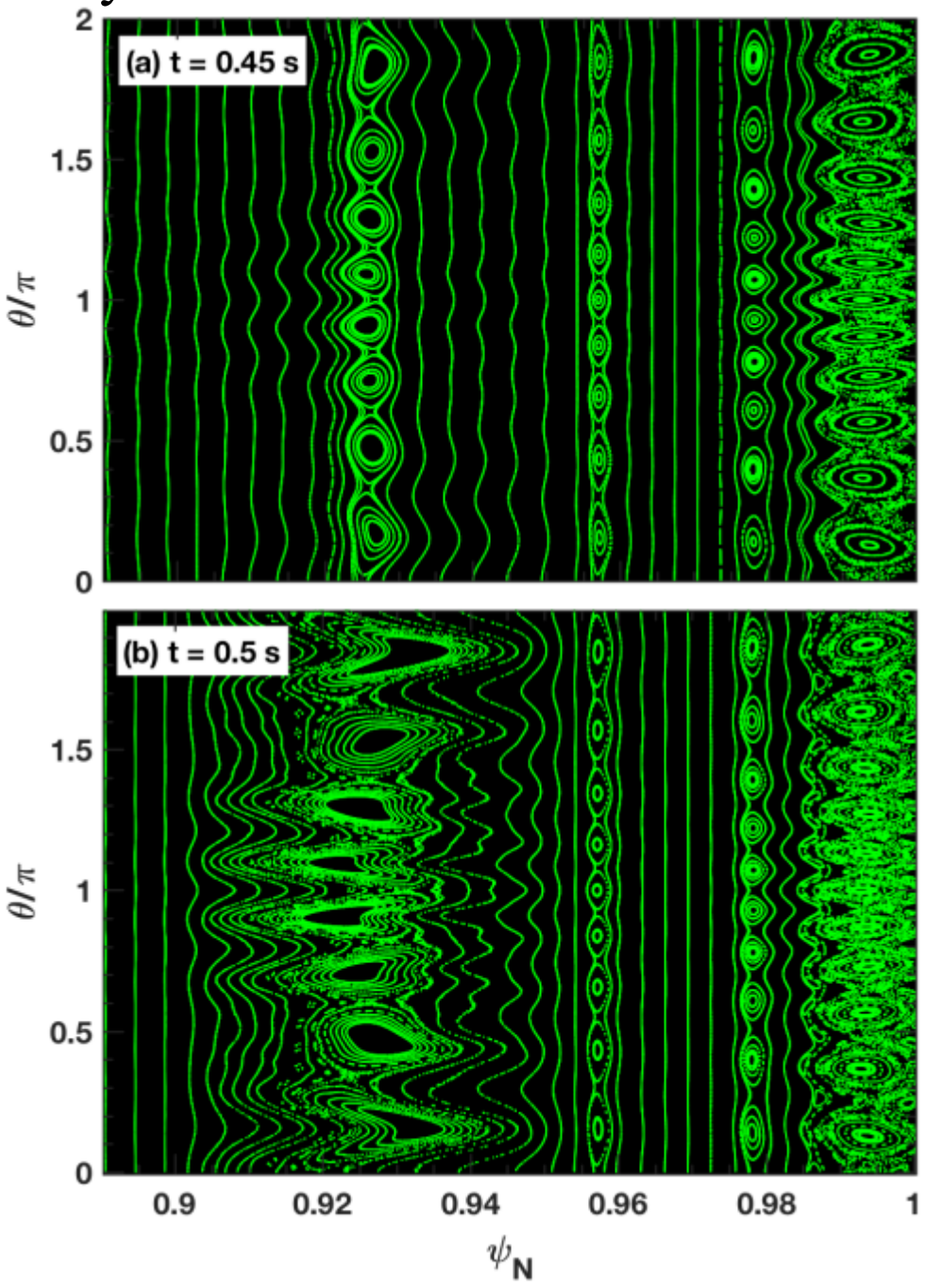

**Figure 5**. Poincaré plot of the poloidal flux surfaces (a) during 11/2 penetration but before 8/2 penetration at 0.45 s in figure 4, and (b) during 11/2 and 8/2 penetration at 0.5 s in figure 4.

Before $m/n$ = 8/2 penetration at the top of the pedestal, there is a narrow island ($\Delta\psi_N \approx 0.015$) chain at the foot of the pedestal (at $q$ = 11/2) corresponding to the $m/n$ = 11/2 penetrated field as shown in figure 5(a). The $m/n$ = 11/2 island penetrates readily at the foot of the pedestal due to the order of magnitude higher local resistivity and lack of perpendicular flows that could screen the 11/2 resonant field at that location, in contrast to the conditions at the top of the pedestal. Strong screening currents are driven at the $q$ = 8/2, 9/2 and 10/2 rational surfaces, lead to only very narrow residual magnetic islands in the gradient region of the pedestal and at the top of the pedestal ($\Delta\psi_N$ < 0.005).

The Poincaré plot of the magnetic surfaces during $m/n$ = 8/2 penetration is shown in figure 5(b). We now see two dominant island chains with island width ($\Delta\psi_N \approx$ 0.01-0.02) at the top ($q$ = 8/2) and at the foot ($q$ = 11/2) of the pedestal. In both cases, strong screening is still obtained in the radial region $\psi_N$ = 0.94-0.98 within the gradient pedestal. The TM1 simulation of screening between the top and foot of the pedestal is consistent with the preservation of the edge temperature gradient. We will shortly see how sensitive this screening is to the resistivity used in the calculation, and hence how sensitive nonlinear MHD simulation is to the choice of input parameters.

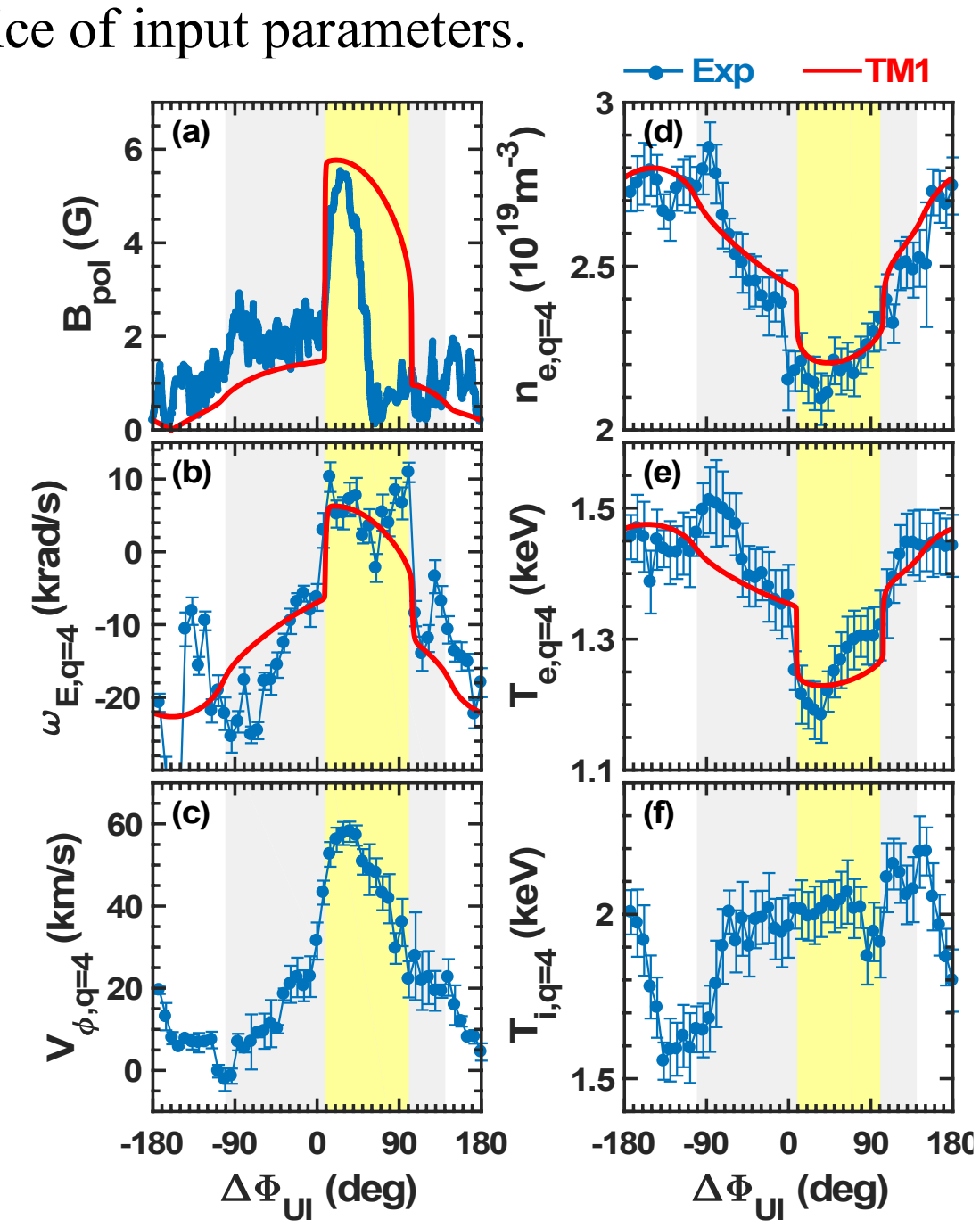

**Figure 6**. Comparison between experiment (blue) and TM1 simulation (red) for (a) the $n$ = 2 poloidal magnetic perturbation on the high-field-side midplane $B_{pol}$, (b) $\omega_E$, (c) experimental carbon-VI toroidal rotation $V_\phi$, (d) $n_e$, (e) $T_e$, and (f) experimental $T_i$ for carbon-VI at the $q$ = 4 rational surface versus $\Delta\Phi_{UI}$.

Figure 6 compares the evolution of the plasma response from TM1 (red) and experiment (blue) for the low-density discharge (#158115) where ELM suppression is observed. The TM1 simulated $B_{pol}$ at $\psi_N$ = 1.1 (figure 6(a), red) shows a jump at the onset of $m/n$ = 8/2 island penetration that is similar in magnitude to the experimental poloidal field jump at roughly the same radius ($\psi_N$ = 1.1) on HFS magnetic sensors (figure 6(a), blue).

The details of the magnetic response between TM1 and experiment differ in duration, but the model captures the magnitude of the observed change, consistent with the change being induced by pedestal-top $m/n$ = 8/2 resonant field penetration. The magnitude and duration of the concomitant jumps in the measured $\omega_E$, $n_e$ and $T_e$ at the $q$ = 4 surface are shown in figure 6(b)-(d) (blue) which are quantitatively consistent with the TM1 simulation (red) at the onset of ELM suppression.

Note that anomalous transport coefficients are inferred from TRANSP analysis of profiles measured before the RMP is applied. TRANSP infers the transport coefficients from power balance analysis including turbulence effects. The TM1 model then computes the effect of additional parallel collisional transport due to magnetic islands. Our analysis does not use profile and transport information during the RMP phase, but rather computes such profiles and transport coefficients for comparison with measurements during the RMP phase. It is well known that the turbulence level is observed to increase following the transition to ELM suppression [15,20], while our analysis takes no account of such turbulence effects on transport. It is therefore quite surprising that TM1 can reproduce the electron response to the RMP at the pedestal top, despite its simplicity.

On the other hand, we also show the carbon-VI toroidal rotation and temperature at the top of the pedestal (figure 6(c) and (f)) that is not modeled in TM1. In this regard we note that TM1 models particle transport, electron thermal transport and momentum transport induced by RMPs. Ion thermal transport and viscosity is not modeled and nor is the toroidal ion velocity, only the total electric field. In addition to these deficiencies in TM1, any effect that leads to enhanced perpendicular transport between rational surfaces is not modeled in TM1. Given these limitations, it is again quite interesting to find such good agreement with the density, electron temperature and radial electric field at the pedestal top.

While the discussion of ion transport is beyond the scope of our paper, we note that the carbon-VI temperature does not respond to the 8/2 penetration at the $q$ = 4 surface (figure 2(g) and figure 6(f)). On the other hand, there is considerable evidence for significant discrepancies between carbon-VI and deuterium temperature profiles at low collisionality in the pedestal [65]. This suggests that the validation of any model of RMP induced ion thermal transport (TM1 excluded) must await further developments in measurement capability.

Importantly, the increase in $\omega_E$ at the $q$ = 4 surface upon penetration of the 8/2 island is derived from carbon-VI measurements, and this measurement is consistent with TM1 predictions. While we lack main ion rotation measurements, we still expect radial force balance to be valid for each ion species, so that the change in $\omega_E$ must be the same for deuterium and carbon-VI. The decrease in the magnitude of $\omega_E$ upon 8/2 penetration reflects the expected reduction in the relative rotation between the plasma at the $q$ = 4 surface and the RMP. An important contributor to the change in $\omega_E$ upon penetration of the $m/n$ = 8/2 island is the increase in co-$I_p$ toroidal rotation of carbon-VI (figure 6(c)).

Figure 7 compares the profiles of $n_e$, $T_e$, $\omega_E$ and perpendicular electron flow frequency $\omega_{\perp e}$ = $\omega_{*e}$+$\omega_E$ from experiment and TM1 simulation for discharge #158115. The initial experimental profiles, taken when the RMP is negligible, are shown in black and are the same as the blue profiles shown in figure 3. The experimental profiles just before ELM suppression are shown in blue dotted lines and the TM1 simulation results are shown in red solid lines in figure 7(a)-(d). The experimental profiles during ELM suppression are shown in blue dotted lines and the TM1 simulation results are shown in red solid lines on figure 7(e)-(h). For the time just before $m/n$ = 8/2 penetration (figure 7(a)-(d)) the simulated profiles of $n_e$ and $T_e$ show flattening at the $q$ = 11/2 rational surface at the foot of the pedestal due to $m/n$ = 11/2 island formation at $\psi_N \approx 0.99$. The TM1 simulation of density flattening at the foot of the pedestal is shown in the inset profiles in figure 7(a)-(b). The simulations also show no penetration in the gradient region of the pedestal or at the top of the pedestal, so the gradients in the density and electron temperature only change at the foot of the pedestal in TM1, producing an experimentally consistent change in the density and temperature at the top of the pedestal.

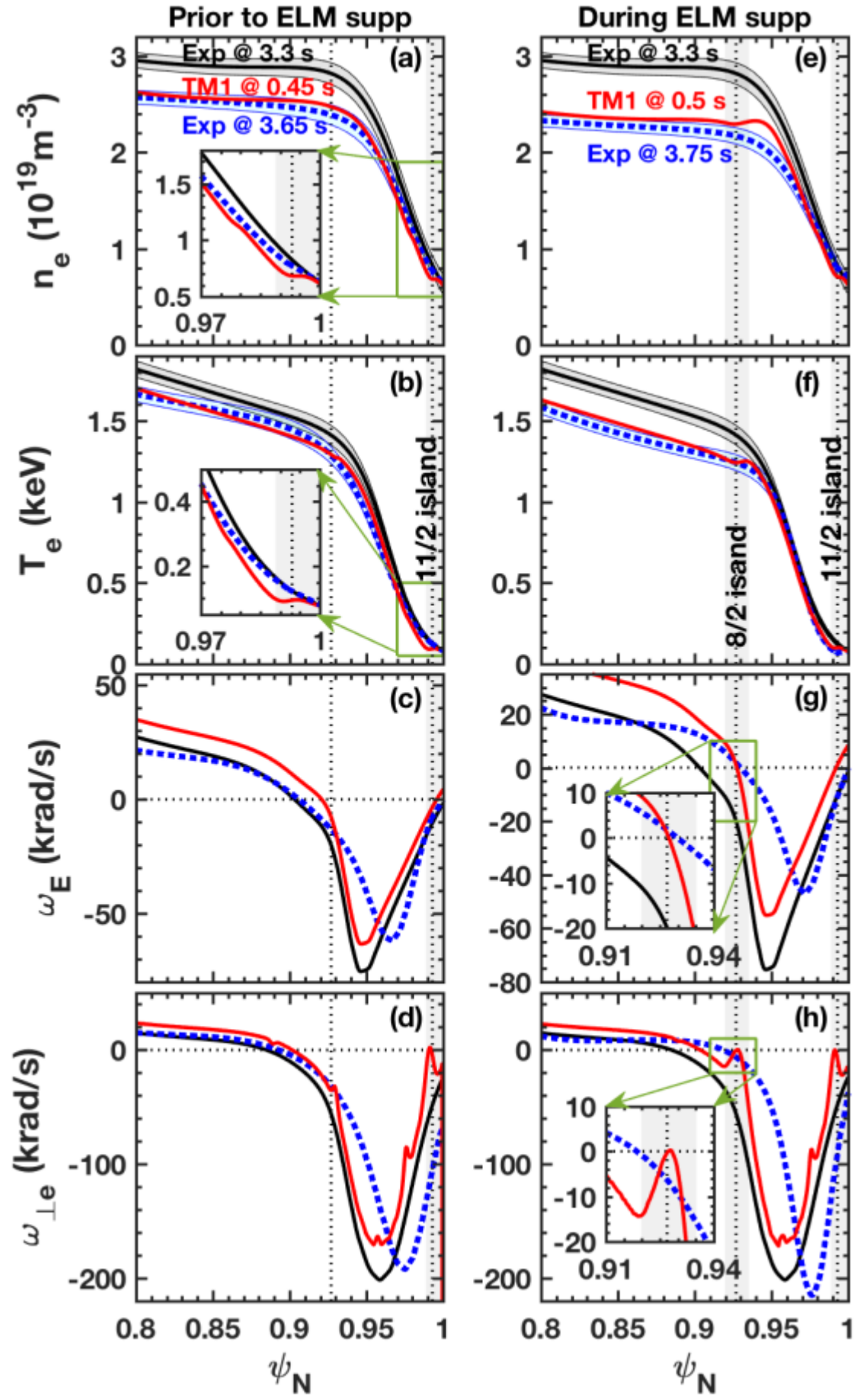

**Figure 7**. Comparison between experiment (dashed blue) and TM1 simulation (red) for the profiles of $n_e$, $T_e$, $\omega_E$ and $\omega_{\perp e}$ (a-d) prior to ELM suppression and (e-h) during ELM suppression for discharge #158115. The initial experimental profiles with minimum RMP are shown in black. The experimental uncertainties in the profiles of $n_e$ and $T_e$ are indicates by the banded blue and red regions in (a), (b), (e), (f).

An enduring mystery in density pump-out is why the density changes more than the temperature at the pedestal top. We resolve this mystery by noting that at the foot of the pedestal, the plasma density has a much shorter scale length than the electron temperature. Consequently, a complete flattening of the electron density and temperature at the foot of the pedestal will only result in a modest decrease in the electron temperature at the pedestal top relative to the change in the density. For a given island width $W$ at the foot of the pedestal, and assuming complete flattening of $n_e$ and $T_e$ over the width of the island, then the relative change at the pedestal top is given by $W\times n'_{e,foot}/n_{e,q=4}$, and $W\times T'_{e,foot}/T_{e,q=4}$ for the density and temperature, respectively. Thus, in figure 8(a) we plot the quantity proportional to the fractional change at the pedestal top. We see straight away that the flattening across the foot of the pedestal has a stronger fractional effect on the density at the pedestal top than the temperature. This basically resolved the long-standing mystery of density pump-out. The curious phenomenology of density pump-out simply arises from the steeper density gradient relative to the temperature gradient near the separatrix. Note that the two curves cross over going from the foot to the top of the pedestal in figure 8(a). This means that a magnetic island at the top of the pedestal will result in a dominant temperature reduction in contrast to the magnetic island at the foot of the pedestal.

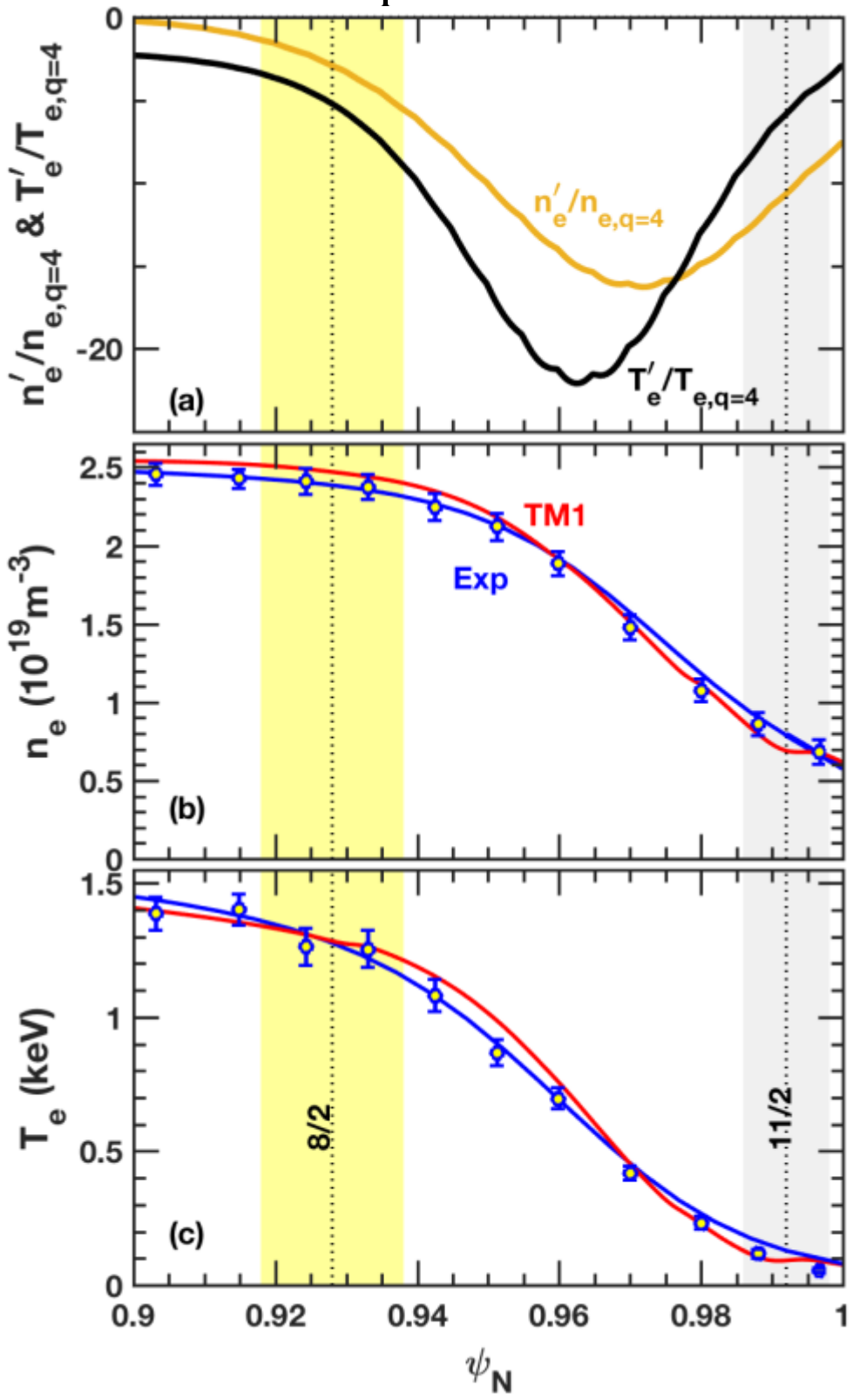

**Figure 8**. (a) Profiles of electron density and temperature gradient normalized by $n_e$ and $T_e$ at the top of the pedestal for #158115 at 3.65 s just prior to ELM suppression. Comparison between Thomson scattering data (blue circles) with fitted profiles (blue solid lines) and TM1 simulated profiles (red) for (b) $n_e$ and (c) $T_e$ during 11/2 penetration.

A key question is whether the local flattening at the foot of the pedestal can be resolved experimentally. Figure 8(b)-(c) show the simulated (red) and measured (blue) density and temperature profiles just prior to ELM suppression. The data points are also displayed (blue symbols). The data shows that the spatial resolution of the Thomson scattering system is inadequate to resolve such narrow (~0.01 $\psi_N$) profile structures. Future studies should aim to resolve these structures by measuring profiles in the flux expansion region near the X-point. Flux compression on the low field side of the magnetic axis makes it particularly difficult to resolve narrow profile structures using conventional diagnostics. Nonetheless, the agreement between TM1 and experiment at the top of the pedestal for pump-out is quite striking. Later we show that the degree of flattening depends on the density through the local resistivity, and thus we can explain how pump-out disappears at high density.

If pump-out results from the flattening across edge magnetic islands, then $m/n$ = 8/2 penetration at the top of the pedestal should also produce pump-out. However, in the case of top of pedestal islands, we see from figure 8(a) that the effect on the temperature must be stronger than the effect on the density because the two curves (green and black) cross over each other between the foot and the top of the pedestal. Indeed, this prediction is confirmed experimentally. We see in figure 6 (yellow band) and in figure 7(e)-(f) that there is a further reduction in the top of pedestal density and temperature during 8/2 penetration. However, the effect on $T_{e,ped}$ (12%) is larger than the effect on $n_{e,ped}$ (8%), unlike the island at the foot of the pedestal. From figure 8(a), we see that the relative effect of flattening across the 8/2 island on $T_e$ and $n_e$ has reversed, compared with the effect of flattening across the 11/2 island.

It must be pointed out that there are profile effects induced by the RMP that simply cannot be reproduced using TM1, as shown in figure 7. The RMP affects the $\omega_E$ profile in ways that are not captured by TM1 except for the local magnitude of the change near the $q$ = 8/2 and 11/2 rational surfaces, as shown in figure 7(d) and 7(h). During 8/2 penetration in figure 7(h) the TM1 simulation shows an upward shift in the $\omega_E$ profile consistent with the reduction in the magnitude of $\omega_E$ at the $q$ = 4 rational surface due to field penetration. This upward shift in $\omega_E$ is due to the **J**×**B** torque exerted by the RMP. However, both in the period during pump-out and during ELM suppression the experimental $\omega_E$ profile (blue dashed lines) shows significant variations from the TM1 prediction (red). This underscores a few important points concerning the limitations of TM1. While the experimental $\omega_E$ matches the TM1 $\omega_E$ at the $q$ = 4 surface during ELM suppression, TM1 does not include non-ambipolar transport mechanisms or other cross-field transport effects induced by the RMP that could affect the $E_r$ evolution [34]. It should also be mentioned that TM1 cannot reproduce the change in the **E**×**B** shearing rate ($\omega_{E\times B} = \omega_E'$) at the top of the pedestal during pump-out and $m/n$ = 8/2 penetration. It is quite likely that these changes in shearing rate are tied to non-ambipolar and turbulent transport mechanisms [21].

Within the two-fluid MHD description, resonant field penetration requires that the electron perpendicular flow to go to zero at the resonant surface, i.e., $\omega_{\perp e} = \omega_{*e} + \omega_E \sim 0$. However, the experimental $\omega_{\perp e}$ is about -7 krad/s at the rational surface from the inset in figure 7(h). More recent systematic analysis of DIII-D data revealed a consistent and significant deviation of $\omega_{\perp e}$ from zero at the rational surface [30]. On the other hand, we find in general that $\omega_E \sim 0$ at the 8/2 rational surface (figure 7(g)). It seems unclear why $\omega_E \sim 0$ in ELM suppression [29,30] whereas the more fundamental quantity $\omega_{\perp e}$ deviates from zero.

This discrepancy in the perpendicular electron flow is not a trivial or unimportant detail. The measured difference between $\omega_E$ and $\omega_{\perp e}$ at the resonant surface is the reason why previous studies used linear single fluid MHD theory to predict the $n$ = 2 linear island size at the $q$ = 4 surface rather than two-fluid linear theory which showed strong screening at the $q$ = 4 surface [15]. In linear single fluid MHD theory, islands form at rational surfaces where $\omega_E \approx 0$ whereas two-fluid linear theory requires $\omega_{\perp e} \approx 0$.

One possible explanation for the discrepancy in the experimental $\omega_{\perp e}$ is that we do not resolve the fine scale structure of the electron pressure in the resonant layer. From figure 8 we see that the Thomson scattering channel spacing is of the order of the island width. If small-scale features arise in $\omega_{*e}$ (as seen in the TM1 simulation in figure 7(h)) then these features will not be captured by the profile diagnostic. From the inset box in figure 7(h), the TM1 simulation shows $\omega_{\perp e} = 0$ in a very narrow region around the rational surface but falls away quickly on either side.

Another question is why the TM1 simulation and experiment show $\omega_E \approx 0$ during ELM suppression. If we suppose that on a fine spatial scale the diamagnetic frequency is $\omega_{*e} \approx 0$, then it must follow that $\omega_E \approx 0$ at the same surface. On the other hand, while fine scale structure may exist in the electron pressure profile, the plasma rotation and ion pressure cannot exhibit such narrow structures due to the much larger ion orbit widths. Thus the structure of $\omega_E$ around the resonant surface may be resolved using existing diagnostics while the fine scale structure of $\omega_{\perp e}$ can remain unresolved, as suggested elsewhere [30]. Future studies need to explore these fine scale structures as part of an effort to identify the magnetic islands from profile measurements. We should also mention that another possibility exists where plasma impurities affect the two-fluid resonance condition [65].

A very important result of our present study with TM1 is that the nonlinear two-fluid MHD model readily recovers the penetration of a magnetic island at the top of the pedestal despite the challenges encountered using two-fluid linear models. Our analysis underscores the importance of using a nonlinear two-fluid theory to predict island formation based on initial conditions rather than to interpret the ELM suppressed profiles for the presence of possible islands using linear models.

Next, an important question is how magnetic island formation at the pedestal top can stabilize Peeling-Ballooning-Modes (PBMs) leading to ELM suppression. From figure 7 we see that the island at the $q$ = 4 surface flattens the density and electron temperature profiles near the top of the pedestal, leading to a reduction in the pedestal height and width consistent with experiment. To show how the magnetic island can limit the pedestal from expanding to an unstable height and width, we plot in figure 9 the measured pedestal height and width (in blue) and compare with the EPED [27] prediction (black) for PBM onset. We note that the experimental width is a much more challenging measurement than the pedestal height, however the data from these experiments is sufficient to resolve both the height and the width. The pedestal electron pressure from experiment (blue circles in figure 9 (a)) shows a drop during periods of ELM suppression (yellow bands), as noted earlier. The TM1 model clearly reproduces the magnitude of the electron pressure reduction during ELM suppression. The drop in the electron pressure is ≈ 20% below the EPED prediction for PBM onset during ELM suppression (yellow bands) from experiment and TM1. A similar magnitude of reduction is seen in the pedestal width. This reduction in the pedestal pressure and width from TM1 is consistent with experiment and indicates that the PBM can be stabilized by the effect of a magnetic island at the top of the pedestal by preventing the pedestal from expanding to an unstable width. Linear stability analysis using the ELITE code also predicts the stabilization of PBMs during ELM suppression [3] due to a contraction in the height and width of the pedestal, consistent with the results in figure 9. Thus, for low-collisionality ISS plasma in DIII-D, the mechanism for ELM suppression by PBM stabilization due to island formation at the pedestal top is consistent with experiment and nonlinear MHD simulation.

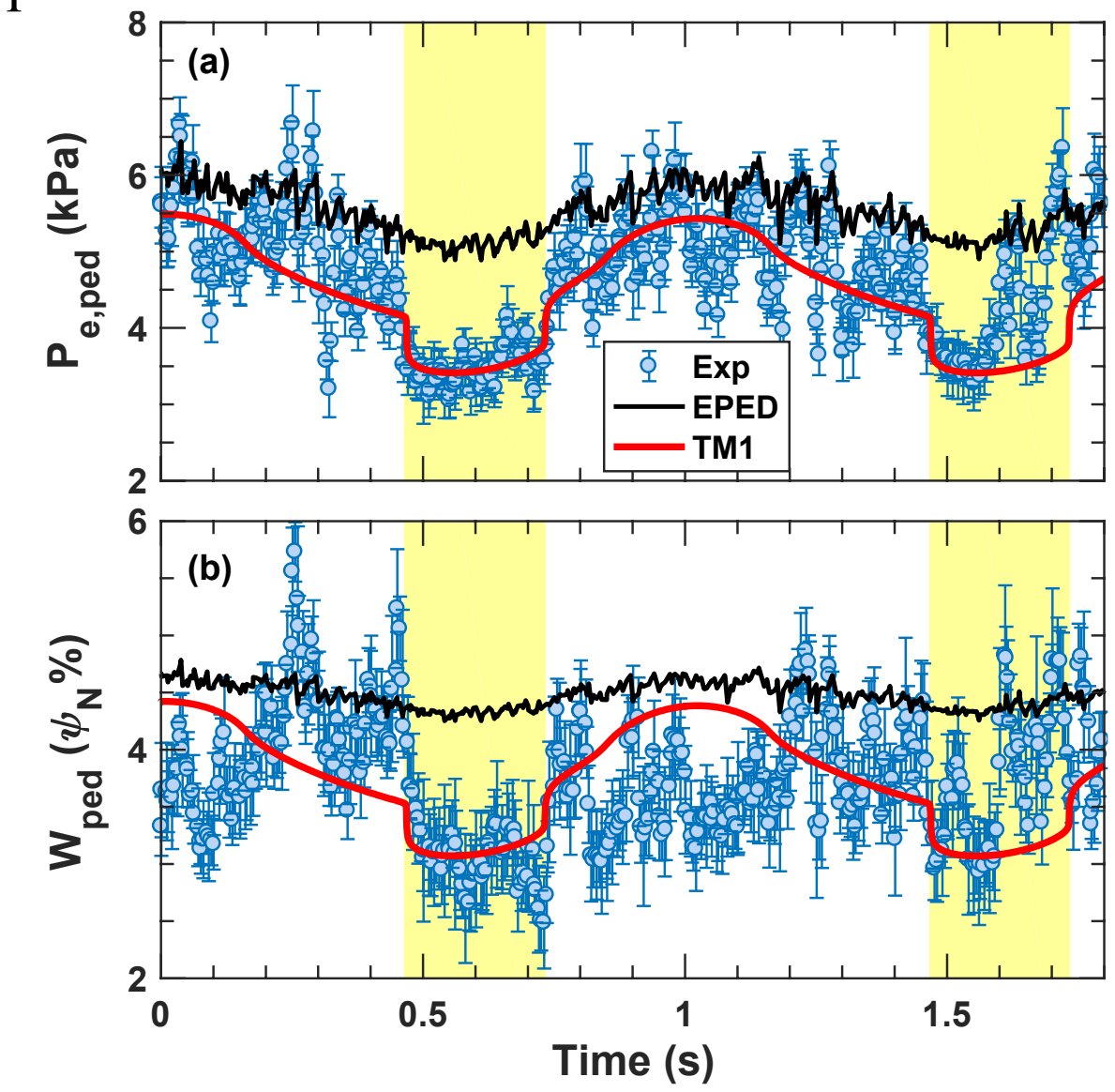

**Figure 9**. Comparison of (a) pedestal pressure $P_{\mathrm{e,ped}}$ and (b) pedestal width $W_{\mathrm{ped}}$ from experiment (blue circles), EPED prediction (black), and TM1 simulation (red) for discharge #158115. Here $t$ = 0 corresponds to $t$ = 3.25 s in figure 2 for the experimental and EPED results.

Our simulations support an early conceptual model for ELM suppression in low-collisionality DIII-D plasmas based on the formation of isolated magnetic islands at the top of the pedestal [27]. So far, we have no evidence from TM1 simulation or experiment for significant stochasticity at the top of the pedestal. Indeed, we also have no evidence for significant field penetration in the gradient region of the pedestal. These results are important because the appearance of significant stochasticity or magnetic islands in the gradient region of the pedestal should result in a very noticeable degradation in the edge temperature gradient, and no such degradation is observed in ELM suppressed plasmas. It is well known from both experiment [66] and theory [10,67] that significant stochasticity in the ETB will collapse the temperature gradient unless nonphysical effects are invoked to prevent such a collapse.

Here we show that the use of a 10X higher plasma resistivity can significantly reduce the threshold for resonant field penetration and lead to greatly enhanced transport due to the generation of islands and stochasticity in the ETB.

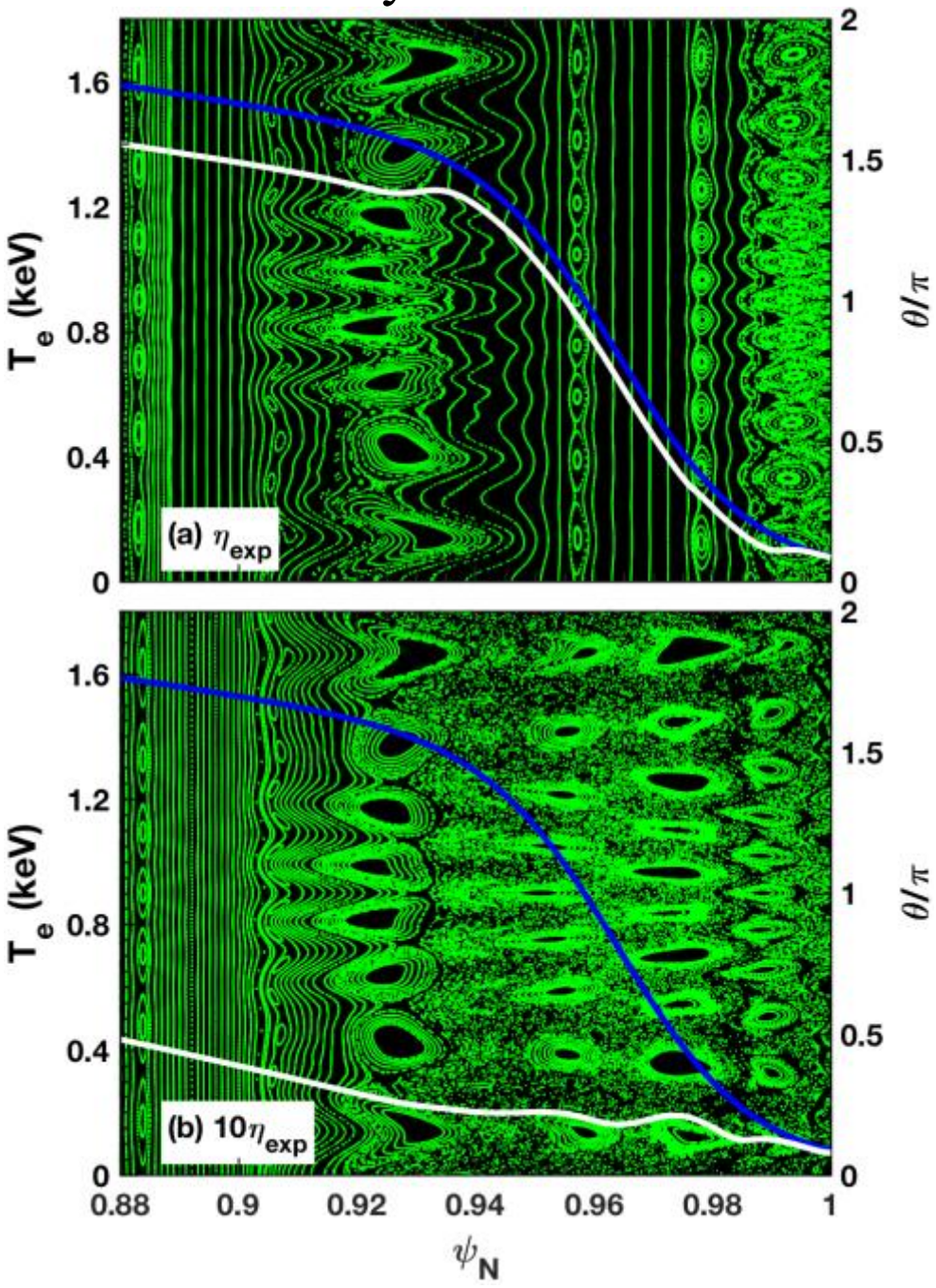

**Figure 10**. The RMP cause wide stochasticity in the pedestal when using 10X higher resistivity than the experimental value. The Poincaré plot of the poloidal flux surfaces is overlaid with the original electron temperature at the minimum of the RMP (blue) and the TM1 simulation (white) for (a) the experimental resistivity and (b) 10X the experimental resistivity for #158115.

TM1 simulation is performed for the low-density discharge #158115 but with 10X the experimental resistivity, and the results are shown in figure 10. Figure 10(a) shows the Poncaré plot of the magnetic field and the simulated electron temperature profile (white) from TM1 for the original profiles and experimental resistivity. The initial temperature profile is shown in blue. In contrast, figure 10(b) shows the TM1 simulation results for the case with identical profiles except for a 10X higher resistivity profile. We now find resonant field penetration throughout the ETB from the foot to the top of the pedestal at every rational surface including the surfaces in the gradient region of the ETB. Not surprisingly this leads to a simulated temperature profile (white) that shows the complete collapse of the ETB to an L-mode like profile. The collapse of the edge temperature profile by multiple magnetic islands has been observed in DIII-D [68] and modeled by TM1 [51] for the case of locked modes leading to a thermal quench. Such a collapse in the pedestal should also be expected with RMPs when using an unrealistically high plasma resistivity. Given that such collapses are inconsistent with ELM suppression experiment, we conclude that it is essential to use the actual plasma resistivity when making predictions on the role of magnetic islands in ELM suppression and density pump-out.

In this section, our nonlinear simulations (figures 4-10) revealed that:

1) The $m/n$ = 11/2 and 8/2 magnetic islands at the bottom and top of the pedestal account for density pump-out and ELM suppression, respectively.
2) The strong screening of resonant fields in the steeper gradient region of the pedestal preserves the ETBs during ELM suppression and pumpout.
3) Increasing the resistivity 10X leads to further field penetration and the collapse of the pedestal in our simulations, highlighting the importance of using experimentally relevant resistivity in nonlinear simulation.

### 3.3. RMP penetration at the pedestal foot and density pump-out

Here we show from TM1 simulation the contribution to density pump-out from different resonant surfaces. Based on figure 7 for the TM1 simulated profiles of the density and temperature just before and during ELM suppression, we show in figure 11 the difference in the density and temperature profiles. Figure 11(a)-(b) show the difference in the TM1 simulated $n_e$ and $T_e$ profiles, respectively, for the time just before 8/2 field penetration (blue) and during 8/2 field penetration (red). The 11/2 field penetration at the pedestal foot is common to both times and strong screening currents are generated at the 10/2 and 9/2 surfaces. These strong screening currents contribute a small amount of pump-out [13,46,47]. However, the bulk of the density reduction at the top of the pedestal comes from the density flattening across the 11/2 island at the foot of the pedestal, as seen in the blue curve in figure 11(a) just prior to 8/2 field penetration. A further reduction in density and temperature occurs at the onset of the 8/2 island, as shown in the red curves. These profiles demonstrate that from TM1 most of the pump-out before ELM suppression comes from a reduction near the foot of the pedestal. As discussed previously, the difference in the reduction in $n_e$ and $T_e$ arise from the difference in the $n_e$ and $T_e$ gradients near the separatrix.

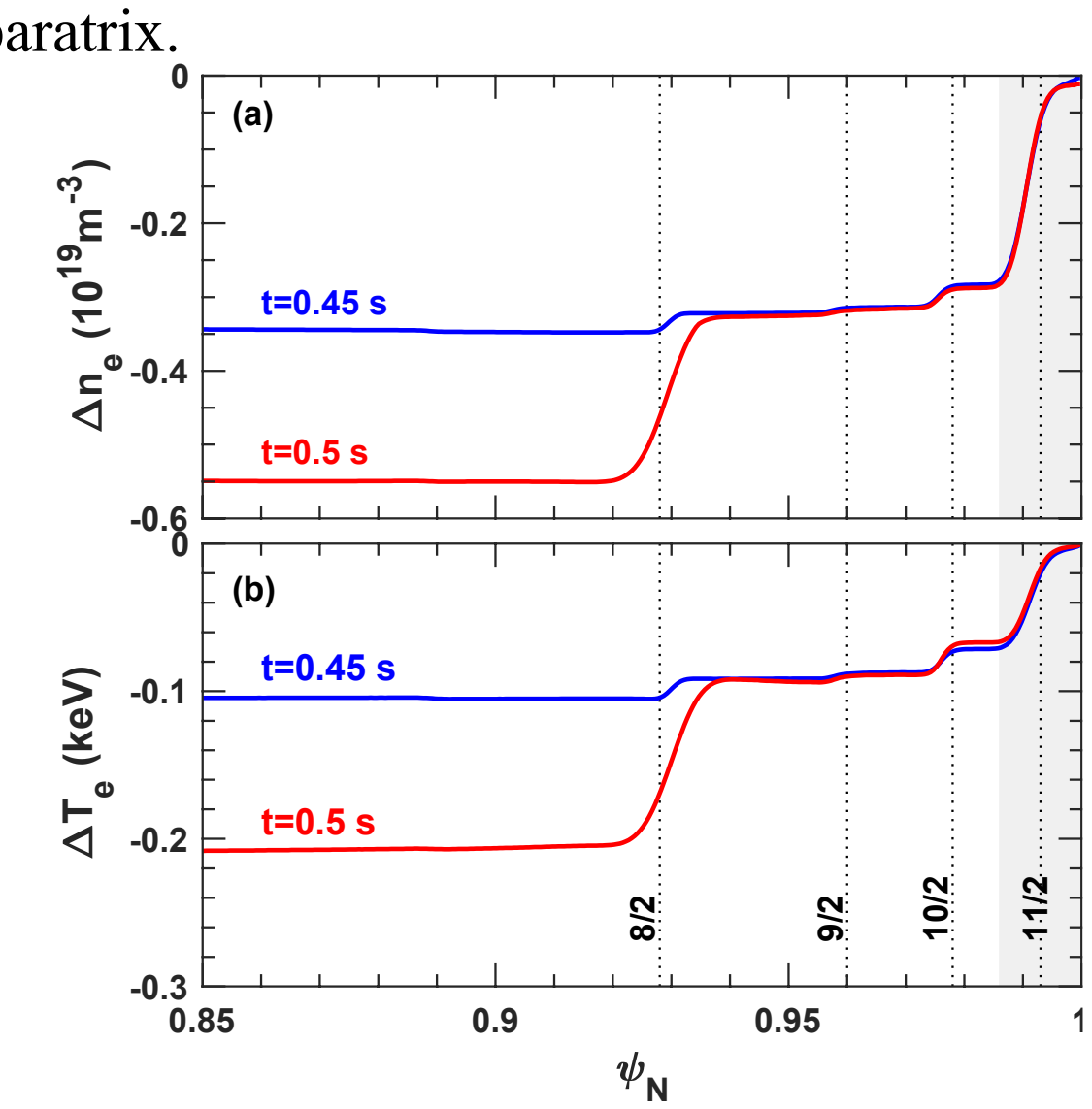

**Figure 11**. The TM1 simulation of the profile change just before 8/2 field penetration (blue) and during 8/2 field penetration (red) for (a) density $\Delta n_e$ and (b) temperature $\Delta T_e$ in discharge #158115.

One important question is why density pump-out decreases with increasing density. Namely, we observe that pump-out is a low collisionality phenomenon [44]. Here we show that as the density increases at the foot of the pedestal, the degree of flattening of the density decreases across the edge islands, leading to a reduction in pump-out. At sufficiently high density, no significant pump-out is observed in TM1 simulations, consistent with experiment.

To further understand density pump-out, simulations are performed for the higher density discharge #159326, shown in figure 2, where ELM suppression is not observed but density pump-out is observed. The initial profiles for these simulations are shown in figure 3 (red curves) when the RMP is negligible. Figure 12 shows TM1 simulation results (red curves) for this discharge including the GPEC calculated sinusoidal variation of the $m/n$ = 7/2, 8/2, 9/2, 10/2 and 11/2 RMP strength. GPEC calculation shows that the peak resonant field strength at the plasma boundary for each harmonic in #159326 is ~10% higher than that of #158115 due to a higher edge pressure.

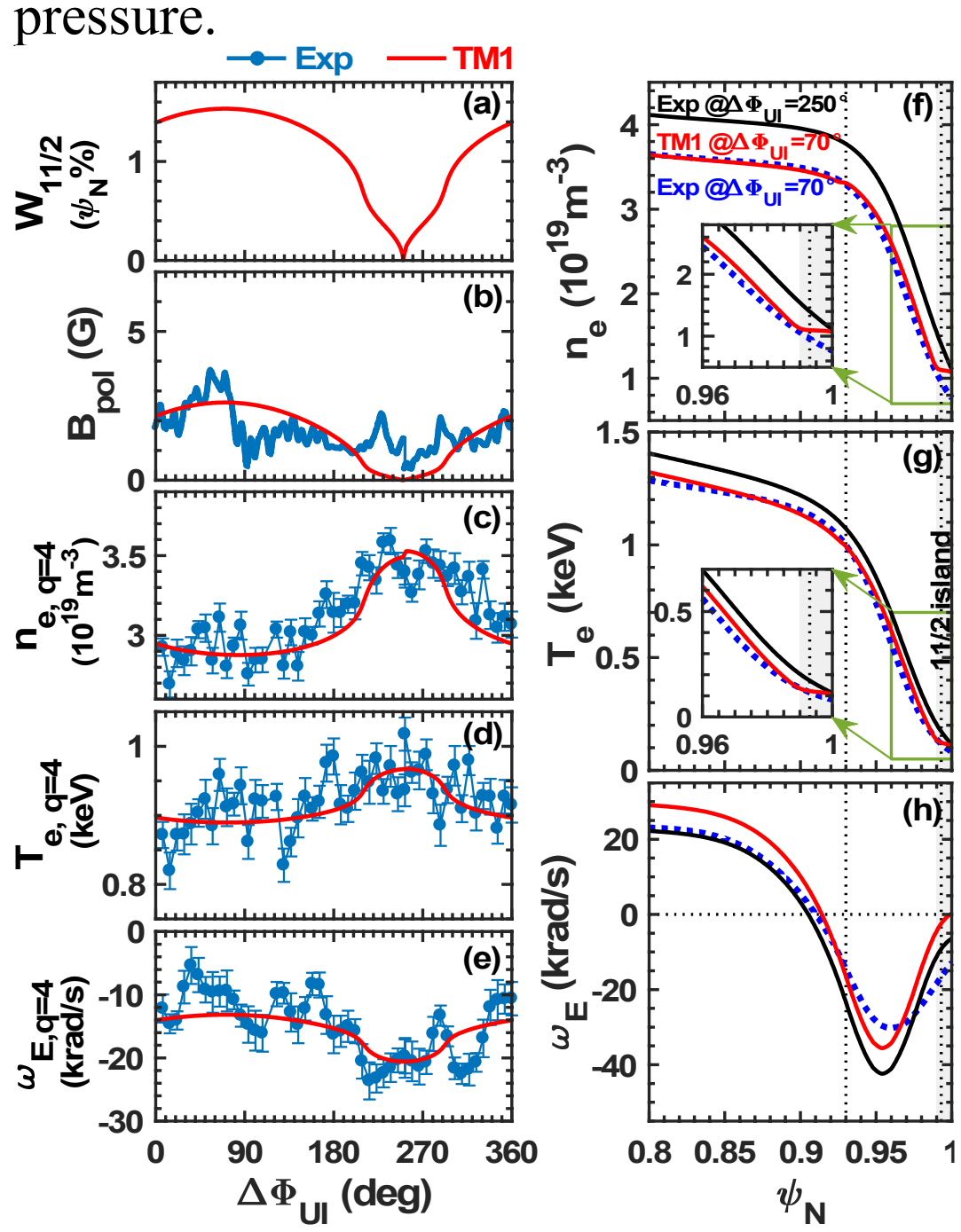

**Figure 12**. Comparison between TM1 simulation (red) and experiment (blue) for (a) magnetic island width $W_{11/2}$, (b) $B_{pol}$ at $\psi_N$ = 1.1, (c) density $n_e$, (d) electron temperature $T_e$, and (e) $\omega_E$ at the $q$ = 4 surface versus $\Delta\Phi_{UL}$. The TM1 simulated profiles (red curves) and experimental profiles (blue curves) are shown for (f) $n_e$, (g) $T_e$, and (h) $\omega_E$ at the maximum RMP strength for discharge #159326 at $\Delta\Phi_{UL}$ = 70 deg. The original profiles (black curves) correspond to $\Delta\Phi_{UL}$ =250 deg where the RMP amplitude is smallest.

The TM1 simulation of pump-out in #159326 reproduces the pump-out observed in experiment, as shown in figure 12(c). The TM1 simulation shows that there is only one significant magnetic island throughout the RMP cycle corresponding to an $m/n$ = 11/2 magnetic island appearing at the foot of the pedestal with a predicted island width of $\approx 0.013\psi_N$ as shown in figure 12(a). The sinusoidal modulation of the 11/2 resonant field causes the modulations seen in the parameters at the top of the pedestal, shown in figure 12(c)-(e), consistent with TM1 simulation.

The TM1 simulated profiles in figure 12(f)-(h) is quantitatively consistent with the experimental profiles. Although the flattening of the profiles at the pedestal foot cannot be resolved in experiment, as discussed previously, the effect seen at the top of the pedestal matches the TM1 simulation. The $\omega_E$ profile is not well reproduced in the TM1 simulation (figure 12(h)) although the discrepancy is not as great as with the lower density profile for discharge #158115. The level of pump-out and the agreement with the experiment profiles of temperature and density is quite clear.

The EPED predicted pedestal width and pressure for ELM onset for discharge #159326 is compared with experiment and TM1 simulation in figure 13. We note from figure 13 that the TM1 simulated pedestal pressure and width do not show any sudden drop and remain close to the EPED prediction and experimental data, consistent with the maintenance of ELMing conditions.

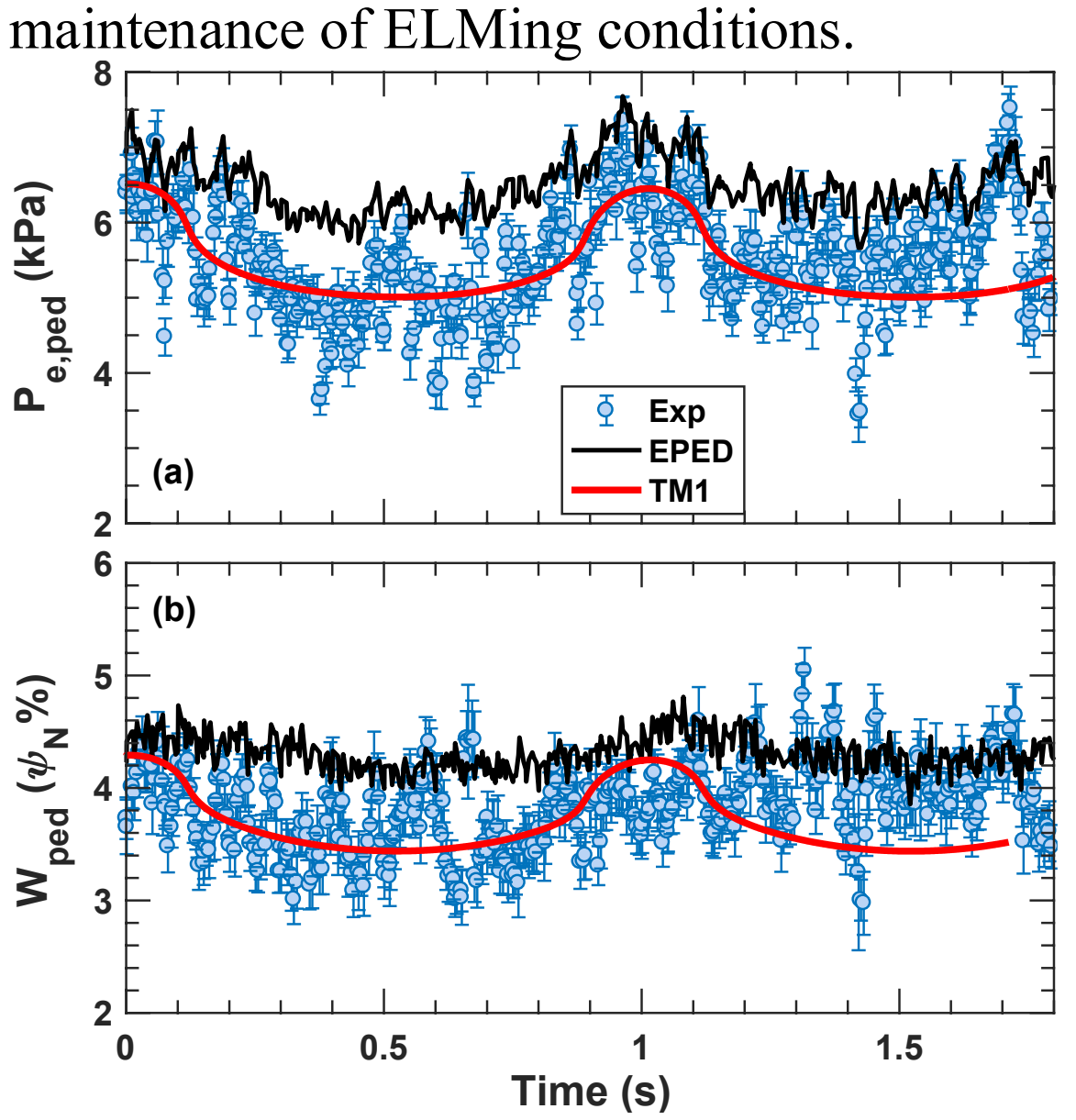

**Figure 13**. Comparison of (a) pedestal pressure $P_{e,ped}$ and (b) pedestal width $W_{ped}$ from experiment (blue circles), EPED prediction of the electron pressure $P_{e,EPED}$ (black), and TM1 simulation (red curves) for discharge #159326. Here $t$ = 0 corresponds to $t$ = 3.4 s in figure 2 for the experimental and EPED results.

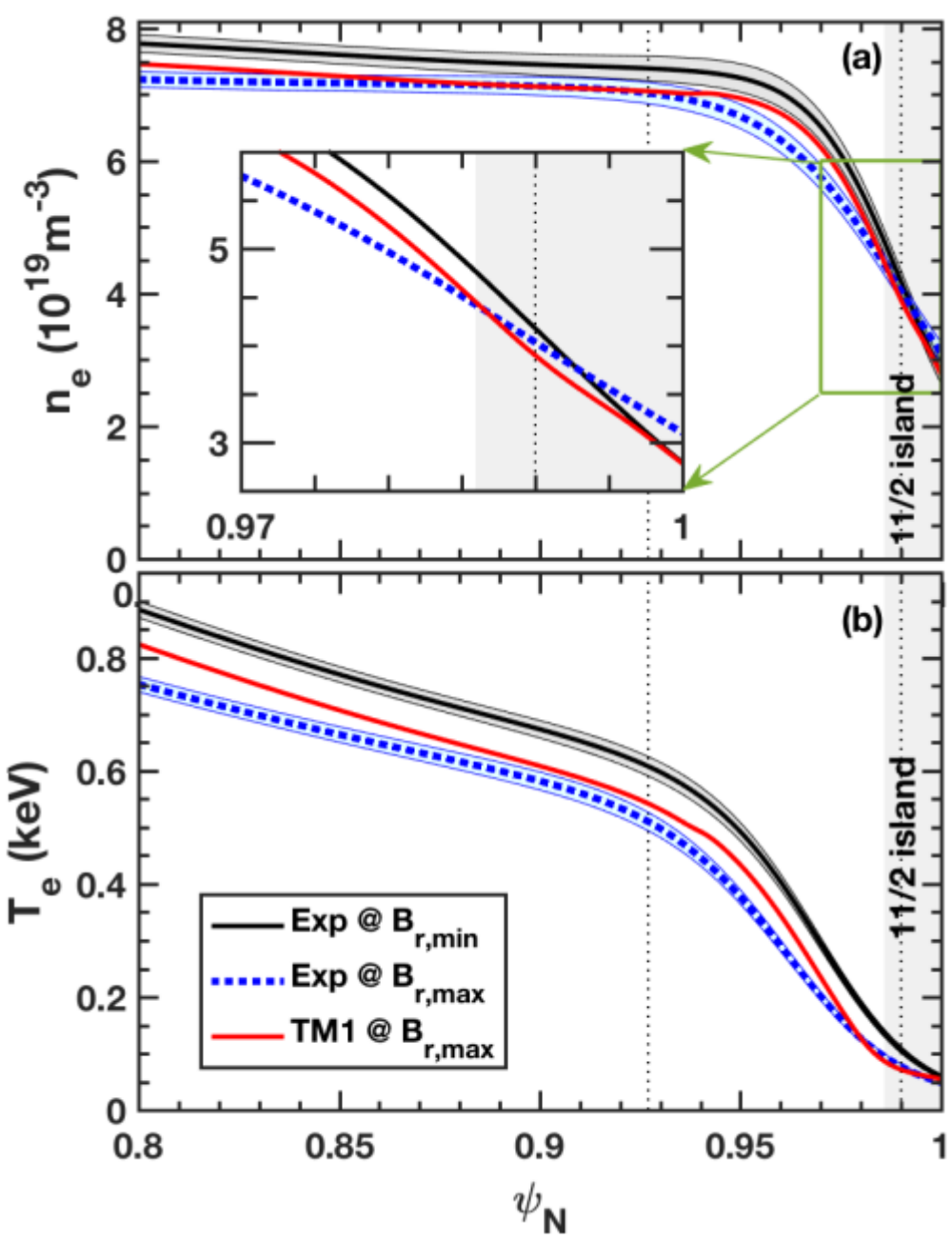

**Figure 14**. Comparison between the TM1 simulation (red curves) and fitting to experimental data (dotted blue curves) for (a) $n_e$ and (b) $T_e$ for the high-density discharge #161202. Here, the black curve corresponding to $B_{r,min}$ is the initial profile for the TM1 simulation at the minimum of the RMP amplitude. The profiles labeled $B_{r,max}$ correspond to the time for the maximum RMP strength when varying $\Delta\Phi_{UL}$.

Next we show that at much higher density the pump-out becomes much weaker from the TM1 simulation. The discharge #161202 shown in figure 14 has a much higher pedestal density of $7.4\times10^{19}$ m$^{-3}$ than the previous discharges, and more than twice that of the intermediate density discharge #159326 shown in figure 12. For #161202, the parameters including $I_p$, $B_t$ and $q_{95}$, shape, and beam power are the same as for #158115 and #159326. The TM1 simulation in figure 14 (red) shows only a small decrease of $n_e$, and a small but comparable decrease in $T_e$, at the full RMP amplitude, consistent with experiment (blue). The initial profile with minimum RMP is in black. Importantly, at high density, there is little or no flattening across the 11/2 island in the TM1 predicted density profile (inset in figure 14(a)), consistent with the lack of significant pump-out observed in experiment.

The analysis presented above shows a trend of decreasing pump-out with increasing density due to the reduced flattening of the density across the islands at the foot of the pedestal. This trend is shown in figure 15, comparing TM1 simulation of pump-out (red) with experiment (blue) over a range of density. Figure 15(a) shows the percentage density change at the top of the pedestal versus the pedestal density when the RMP amplitude is negligible. The decay of the pump-out with increasing plasma density, which has previously been seen in experimental studies [2,3,6], is now reproduced using TM1. We also observe a weaker trend in the pedestal temperature change with density shown in figure 15(b). The simulations are in good agreement with the experimental data, however there is a systematic underestimation of the density from TM1 that may arise from other second order transport effects.

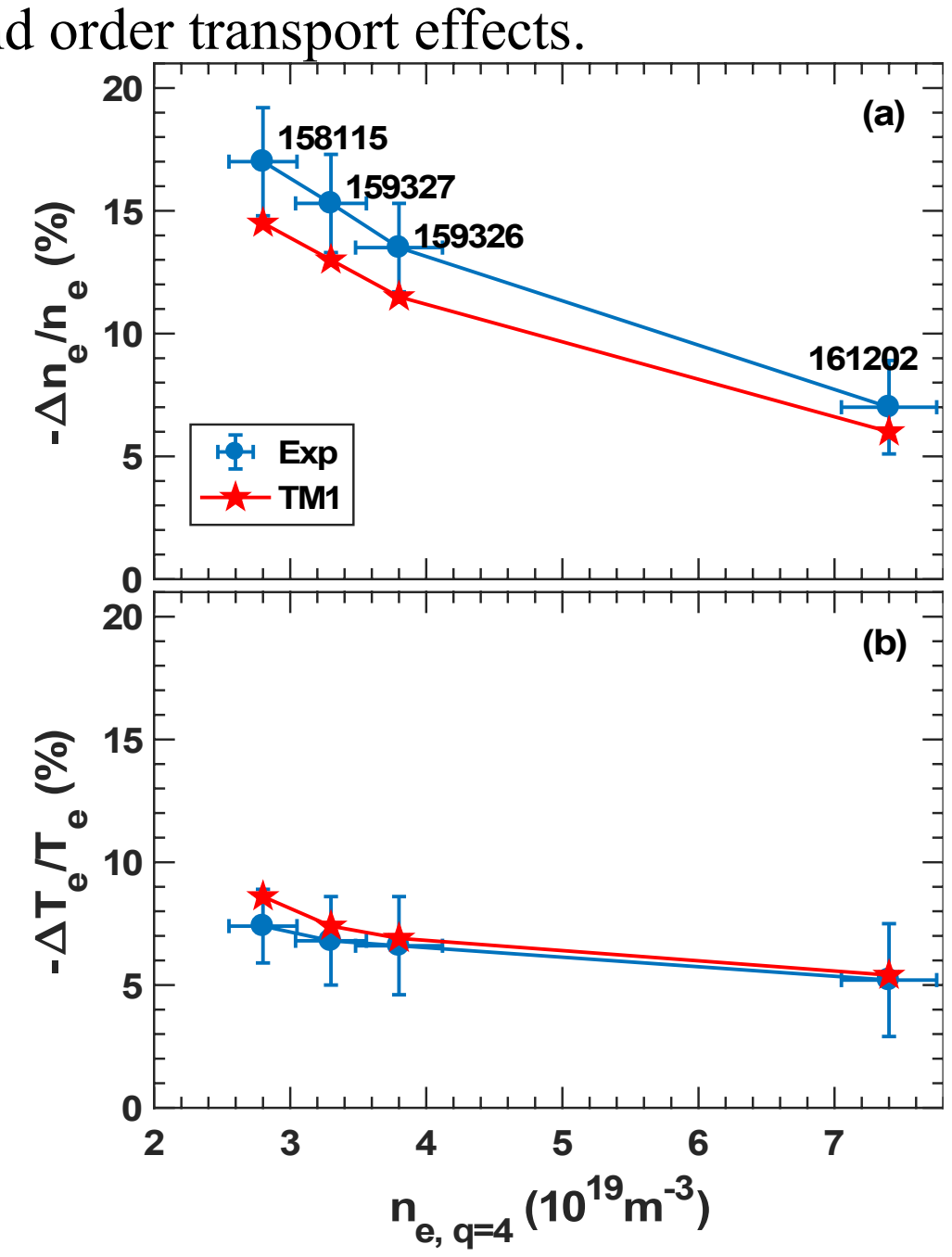

**Figure 15**. Comparison of (a) the density pump-out magnitude $-\Delta n_e/n_e$ and (b) $-\Delta T_e/T_e$ versus density at the $q$ = 4 surface for experiment (blue) and TM1 simulations (red).

An important question is why density flattening across edge magnetic island depends on the plasma density. An approximation using quasi-linear arguments supported by TM1 simulation and theory was presented elsewhere showing that the relative change of density at the rational surface follows [13,46,47]

$$\frac{\Delta n_e}{n_e} \propto \frac{W n_e'}{\nu_{ei}}. \quad (1)$$

Here, $\Delta n_e$ is the change in density, $W$ is the island width, $n_e'$ is the density gradient at the rational surface and $\nu_{ei}$ is the local collisionality. Equation (1) indicates that density pump-out produced by magnetic islands will be weaker at higher collisionality. The reason is that the magnitude of pump-out is proportional to the ion polarization current, which is reversely proportional to plasma collisionality [13,46].

The TM1 simulations of density pump-out and resonant field penetration underscore the critical importance of using the experimentally correct collisionality (or resistivity, since $\eta \propto \nu_{ei}/n_e$) profile in nonlinear MHD simulation. Equation (1) informs that the magnitude of density pump-out will be much weaker if we artificially increase the plasma collisionality by one or two orders of magnitude. As the collisionality increases the flattening at the foot of the pedestal decreases, resulting in less pump-out, as expected from both semi-collisional MHD theory [13], two-fluid [46,47] and gyrokinetic simulation [69].

The model of density pump-out presented here, based on island formation at the pedestal foot, can be used to make quantitative predictions for ITER almost immediately. We know that island formation near the separatrix will occur at low RMP amplitude in ITER because of the weak flow and collisional nature of the plasma near the separatrix. For ITER, expectations are that $T_{e,sep} \approx$ 250 eV and $n_{e,sep} \approx 4\times10^{19}$ m$^{-3}$, leading to a resistivity of about $1\times10^{-6}$ Ωm, which is comparable to the resistivity at the foot of DIII-D plasmas at low density, such as in #158115. Also there are arguments made that the density could be flatter near the separatrix in ITER than in present experiments due to neutral opacity [70]. If the density gradient near the separatrix is significantly weaker in ITER than in present experiments, then density pump-out in ITER may be quite weak compared to DIII-D. This is an important consideration for ITER as a high level of pump-out would place a burden on the tritium recovery system.

A key topic for further study is if pump-out deviates from the TM1 simulations at even lower pedestal collisionality than studied in this paper. There are indicates that at much lower collisionality plasma turbulence [36] may play an increasing role in pump-out. Quantifying the pump-out at lower plasma collisionality with TM1 will be important to help identify gaps in the TM1 model that can be addressed with higher physics fidelity models.

In this section (figures 11-15), we demonstrate from nonlinear simulation that the observed decrease of pumpout with increasing density is quantitatively consistent with the effect of collisionality on parallel collisional transport at the foot of the pedestal.

### 3.4. Prediction for the RMP penetration threshold at the pedestal top

In RMP ELM control experiments in DIII-D, an upper limit of the pedestal density and a lower limit of the pedestal toroidal rotation are observed for accessing ELM suppression [30,44]. Experimentally, in these DIII-D ISS plasmas, if the co-$I_p$ neutral beam torque is decreased below 3 Nm (from typical values of 5-7 Nm) and/or if the density is increased above $\approx 3\times10^{19}$ m$^{-3}$, then ELM suppression cannot be achieved with the available RMP amplitude using the I-coils [30,44].

Motivated by these experimental realities, we undertook to determine if TM1 can reproduce these limits for the ELM suppression threshold. We therefore performed a comprehensive set of simulations of 8/2 field penetration by scanning the $n_e$ and $\omega_E$, profiles up and down in discharge #158115, taking as our reference the profiles just before ELM suppression. Figure 16 shows a contour plot of the threshold $n$ = 2 field $B_{r,th}/B_T$ for resonant field penetration versus $n_e$ and $\omega_E$ at the $q$ = 4 surface. The TM1 prediction of the threshold for 8/2 field penetration is then compared with a database of $n$ = 2 RMP discharges in DIII-D in (yellow stars) and out of ELM suppression (black stars) in figure 16.

The 2D color contours of the $m/n$ = 8/2 penetration threshold $B_{r,th}/B_T$ is shown in figure 16. Here, $B_{r,th}$ is the TM1 predicted resonant field amplitude on the simulation boundary $\psi_N$ = 1 required for resonant field penetration at the top of pedestal. It is found that the required RMP amplitude to penetrate at the top of the pedestal increases for higher pedestal density or for more negative $\omega_E$. This is consistent with the phenomenology observed in DIII-D [30,44]. In the absence of rotation the $\omega_E$ at the top of the pedestal is negative (from the radial pressure gradient). Co-$I_p$ neutral beam injection increases $\omega_E$, making penetration more likely, again consistent with experimental observation [30]. Therefore, as the magnitude of $\omega_{E,q=4}$ decreases for a given density (take any upward pointing vertical line in figure 16), then the threshold for 8/2 field penetration decreases (shown by the color contours going from yellow to blue). On the other hand, for a given $\omega_E$, at the $q$ = 4 surface and by increasing pedestal density (any horizontal right pointing arrow in figure 16), the plasma inertia ($n_e\times\omega_E$) will increase and more **J**×**B** torque will be required for resonant field penetration. In this case we can see that the threshold for field penetration goes from blue to yellow indicating higher RMP amplitude is required for resonant field penetration at higher density. Thus, the trends from TM1 are in qualitative agreement with experiment.

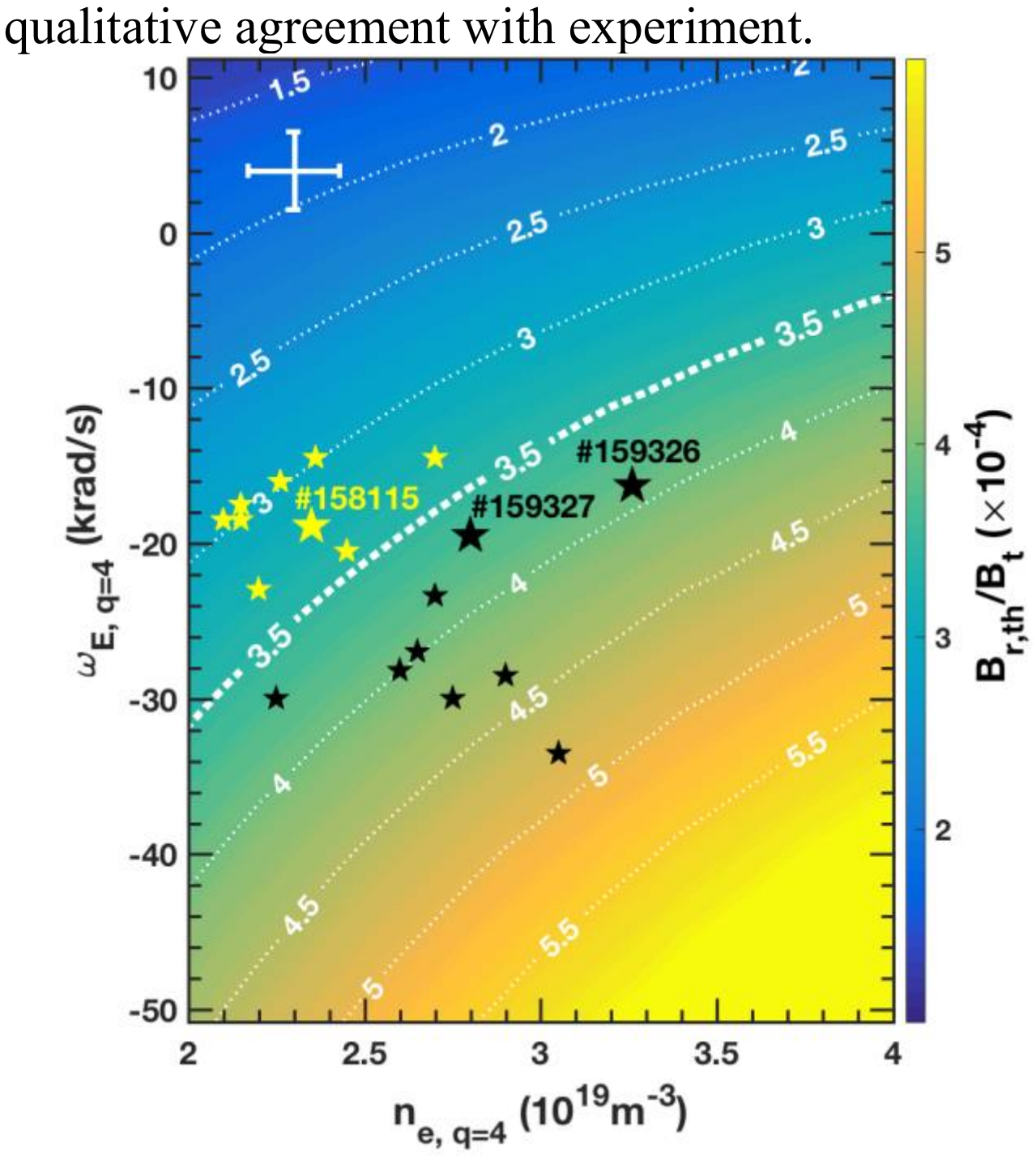

**Figure 16**. Prediction for the RMP penetration threshold at the top of pedestal using the profiles in #158115 just prior to ELM suppression. Contour plot of the 8/2 penetration threshold $B_{r,th}$ from TM1 are shown in the 2D domain of $n_{e,q=4}$ and $\omega_{E,q=4}$ are shown. Nine shots with ELM suppression (yellow stars) and nine shots without ELM suppression (black stars) are overlaid by using the corresponding $n_{e,q=4}$ and $\omega_{E,q=4}$ for each profile. The parameter uncertainty of the database is shown by the cross at the upper left.

Next, we overlay the database of $n$ = 2 RMP discharges with similar ISS conditions and electron temperature in figure 16. The $n_e$ and $\omega_E$ at the $q$ = 4 surface for each discharge is taken just before ELM suppression (yellow stars) and at the maximum of the applied RMP when ELM suppression is not obtained (black stars). We find that the discharges with and without ELM suppression are clearly separated by the RMP contour corresponding to the maximum applied RMP amplitude in these experiments, $B_r/B_T \sim 3.5\times10^{-4}$. ELM suppression is consistently observed for those discharges where the TM1 predicted threshold is lower than the maximum applied RMP amplitude, to the left of the dashed line. Similarly, ELM suppression is not observed in those discharges where the TM1 predicted threshold is above the maximum applied RMP amplitude in these discharges. Therefore, using experimental profiles, actual plasma resistivity and transport coefficients from TRANSP, we recover quantitatively the threshold condition observed for ELM suppression in DIII-D for both the pedestal density and plasma rotation.

Based on the 2D threshold scan we then derive a dimensional scaling relation for the penetration threshold in DIII-D as follows

$$B_{r,th}^{scale}/B_t = 3.5\times10^{-2} n_e^{0.7} |\omega_E + \omega_{*e}|^{0.9} B_t^{-1}. \quad (2)$$

Here all the parameters represent values at the $q = 4$ surface near the pedestal top just before the onset of ELM suppression. The density is in units of $10^{19}$ m$^{-3}$, the rotation frequencies are expressed in units of krad/s and the toroidal field is in Gauss.

The density scaling in Equation (2) is close to analytical predictions for the density dependence of core error field penetration (the coefficient was 2/3 from theory) [71], and the scaling on the rotation frequency is also close to the linear dependence seen for core field penetration in experiment [50,72,73] and theory [49,71]. The only difference between Equation (2) and the theory for core field penetration is that in the pedestal the flow frequency is dominated by the diamagnetic drift frequency ($\omega_{*e} \approx$ -40 krad/s) whereas in the plasma core $\omega_E$ is dominant.

The scaling in Equation (2) reveals two features required to suppress ELM by resonant field penetration in DIII-D. First, at higher density stronger RMP is required for a given $\omega_E$ to trigger field penetration at the pedestal-top and suppress ELM due to the increase in inertia. At the maximum RMP strength available in the experiment, the pedestal density must stay below a critical value to achieve ELM suppression. Such a critical density is observed in DIII-D [30,44] and ASDEX-Upgrade [29]. Second, stronger negative $\omega_E$ (at lower co-$I_p$ neutral beam injection and/or for stronger pressure gradients) also requires higher RMP amplitude to trigger field penetration and suppress ELM. As a result the pedestal-top toroidal rotation must be higher than some threshold to suppress ELM, which is also observed in DIII-D [30].

The qualitative consistency between the scaling law in Equation (2) and experimental conditions required to suppress ELMs in DIII-D strengthens to case that resonant field penetration at the pedestal-top is responsible for ELM suppression. In addition, we now have a scaling relation that allows quantitative predictions for the onset of field penetration in DIII-D and perhaps also for other experiments including ITER.

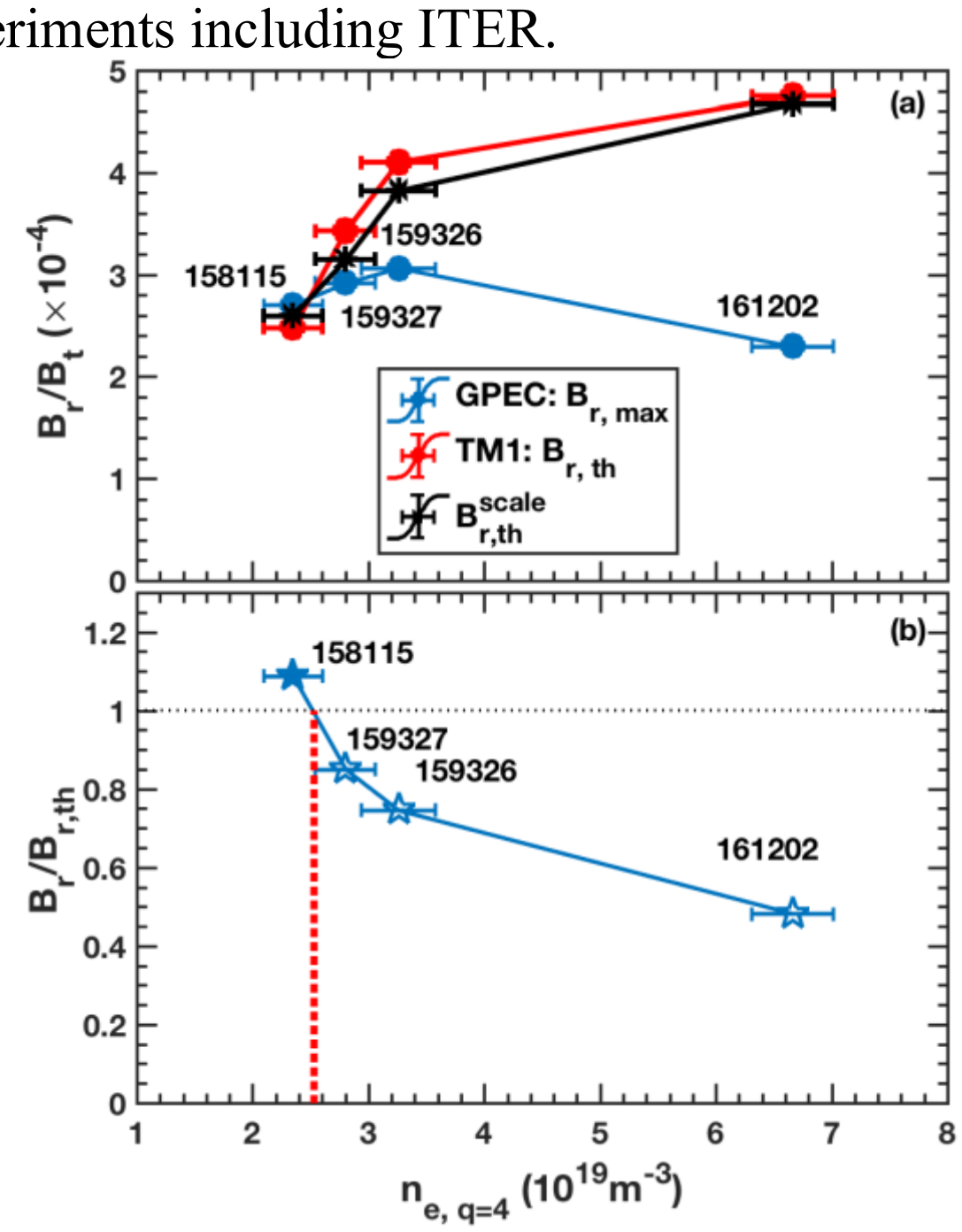

**Figure 17**. (a) The normalized amplitude of plasma response $B_r/B_T$ for $m/n$ = 8/2 (blue circles) calculated by GPEC, the TM1 simulated threshold for 8/2 penetration $B_{r,th}/B_T$ (red circles), and the estimated threshold $B_{r,th}^{scale}$ (black) from Equation (2) are shown as a function of plasma density at the $q$ = 4 surface for shots 158115, 159326, 159327 and 161202. (b) The ratio of $B_r$ and $B_{r,th}$ for these 4 discharges.

Figure 17(a) shows the 8/2 penetration threshold $B_{r,th}$ from TM1 simulation (red), the RMP amplitude $B_r$ from GPEC calculation (blue), and the threshold $B_{r,th}^{scale}$ based on Equation (2) (black) versus pedestal density for discharges 158115, 159326, 159327 and 161202 with $\omega_{\perp e,q=4} \approx$ -40 krad/s. The values of $B_{r,th}$ and $B_r$ are given at the simulation boundary $\psi_N = 1$. $B_{r,th}$ is the field required for penetration according to TM1. $B_r$ is the field calculated from GPEC on the simulation boundary. Only the lowest density plasma exhibits ELM suppression. For all the others, the TM1 inferred threshold is consistent with the scaling and is well above the experimentally available RMP level obtained from GPEC calculation. Figure 17(b) shows the ratio of the experimental RMP amplitude of the 8/2 resonant field to the TM1 simulated 8/2 penetration threshold for the same discharges. A threshold pedestal density is predicted above which ELM suppression does not occur. This threshold pedestal density is $\approx 2.5\times10^{19}$ m$^{-3}$, which is very close to the empirical threshold observed in a variety of DIII-D experiments with $n = 2$ [44] and $n = 3$ RMP [30].

Armed with these predictive results, we proceed to address the penetration threshold for $n = 3$ RMPs at the top of the ITER pedestal. This is necessarily speculative however we have good estimates for the pedestal pressure and density in ITER relevant to $Q_{DT} = 10$ operation and we can therefore assess the consistency of these profiles with resonant field penetration.

**Figure 18**. ITER equilibrium profiles of (a) safety factor $q$, (b) **E**×**B** rotation $\omega_E$ and $\omega_{*e}$, (c) electron density $n_e$ and (d) temperature $T_e$ The $q = m/n$ = 8/3, 9/3, 10/3 and 11/3 rational surfaces are shown by the dotted lines from left to right, respectively.

We consider $n = 3$ RMPs in high power ITER plasmas using equilibrium profiles [74] shown in figure 18. Note that we assume the contribution to $E_r$ from toroidal rotation at the pedestal top is negligible in ITER. This may be an incorrect assumption, however, we now only consider this particular case and defer a thorough parameter scan including rotation to a future work. In addition, we assume that the scale length of the density and temperature at the pedestal top will also be about 10 times smaller in ITER than in present devices based on TGLF transport predictions [61]. Note that this implies a small diamagnetic flow at the top of the pedestal. Here we do not consider resonant field penetration at the foot of the pedestal. To do so we would need a more accurate model of the ITER profiles near the separatrix, and this is also an area of active research [75].

We compare the scaling from Equation (2) with the TM1 simulation threshold for the DIII-D discharges in figure 17 and for the ITER calculation of the $m/n$ = 9/3 threshold for the model profiles in figure 18. Figure 19 shows that the penetration threshold from TM1 (vertical axis) versus the threshold obtained from the scaling (Equation (2) – on the horizontal axis) for the $m/n$ = 9/3 island at the pedestal top in ITER. The TM1 simulated threshold ($B_r/B_t \approx 2\times10^{-5}$) is an order of magnitude lower than for the threshold in DIII-D ISS plasmas and is in good agreement with the scaling in Equation (2).

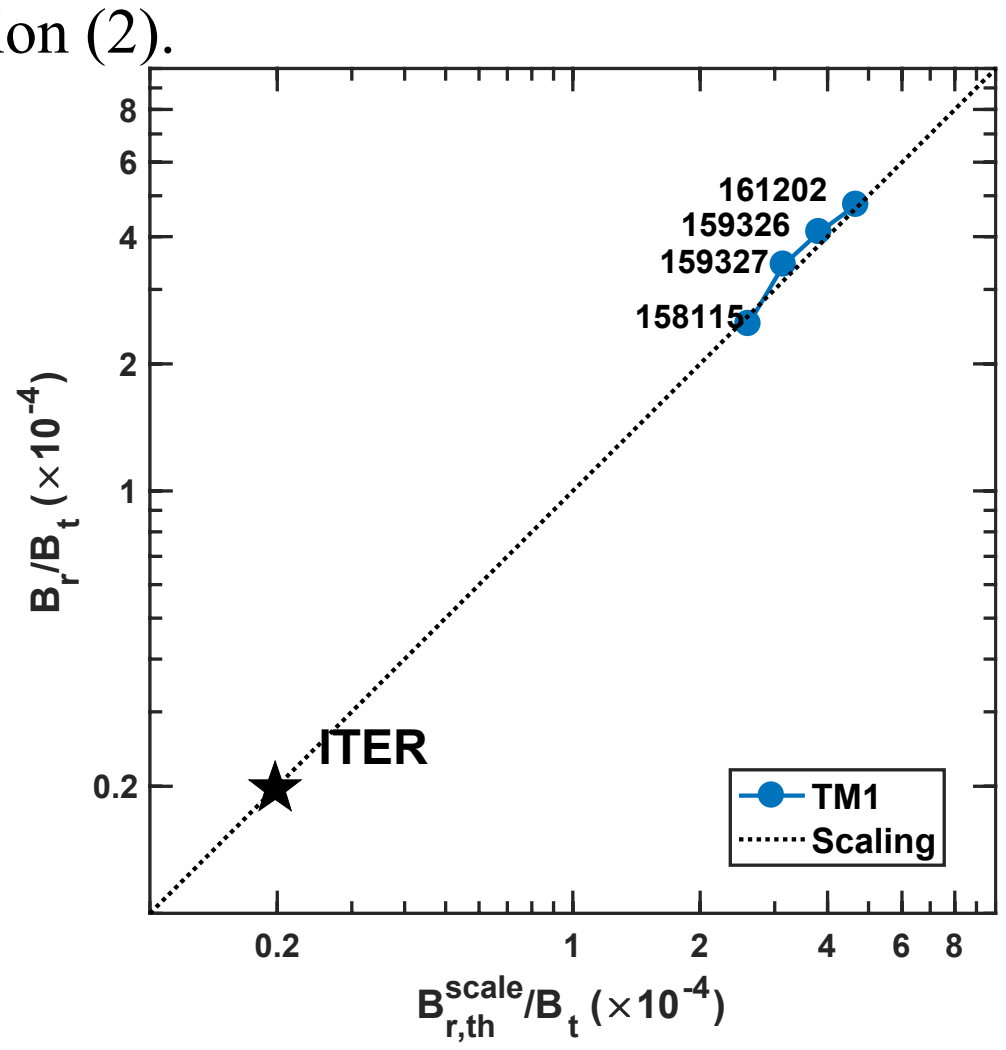

**Figure 19**. $B_{r,th}$ from TM1 simulation (blue circles) are shown as a function of $B^{scale}_{r,th}$ for these 4 shots. The prediction for ITER $n$ = 3 penetration at the top of pedestal according to Equation (2) and TM1simulation is shown together by black five-pointed star.

Note that the resistivity at the top of pedestal in DIII-D ISS plasmas is $\approx 1\times10^{-7}$ Ωm while in ITER it is expected to be $\approx 2\times10^{-8}$ Ωm. The lower resistivity should contribute to stronger screening, but the much lower diamagnetic frequency is the deciding factor for field penetration. For DIII-D we have $\omega_{*e} \approx$ -40 krad/s at the $q$ = 4 surface, whereas we expect $\omega_{*e} \approx$ -3 krad/s at the $q$ = 3 surface at the top of the ITER pedestal [76].

We note that the low value of field penetration in the ITER pedestal is based on the assumption of negligible toroidal rotation contribution to $E_r$. Also, it is not a given that low field penetration will produce ELM suppression at low RMP amplitude in ITER. From DIII-D experiment, we typically observe a pedestal pressure/width reduction of ≈15% below the EPED prediction in ELM suppressed plasmas. Achieving such a pressure and width reduction may require comparable RMP amplitude in ITER. Therefore, the threshold for field penetration and ELM suppression may be quite different in ITER and should be a subject for future study. Nonetheless, a key point is that field penetration should not be difficult to achieve in ITER compared to present day experiments provided the toroidal rotation is sufficiently small in ITER.

In this section, the dependence of ELM suppression on density and **E**×**B** rotation is consistent with the scaling for top of pedestal field penetration based on hundreds of MHD simulations (equation (2)). This relation resembles analytic formulae for core locked modes, but includes the dominant role played by diamagnetic rotation in the pedestal. This derived scaling also matches the $n$ = 3 penetration threshold calculated by TM1 for a set of ITER profile predictions.

### 3.5. Hysteresis effect in ELM suppression

Both in experiment (figure 2) and simulation (figure 4) the RMP amplitude required to produce ELM suppression is higher than the level for transitioning out of ELM suppression, which is a hysteresis effect [77] and frequently observed in experiment for resonant field penetration. The evolution of the TM1 predicted $\omega_E$ at the $q$ = 4 surface is shown as a function of the 8/2 RMP amplitude in figure 20 (red), and the experimental $\omega_E$ at the $q$ = 4 surface is shown in blue. The magnitude of the simulated hysteresis effect is qualitatively consistent with experiment. The evolution of $\omega_E$ behaves as a counterclockwise loop versus $B_r$. This hysteresis effect is due to the balance between the **J**×**B** force caused by the RMP and viscous force [78] in field penetration dynamics. The TM1 simulation of the hysteresis loop is qualitatively consistent with the experimental data, however significant differences arise in the magnitude of the change in $\omega_E$ that requires further study.

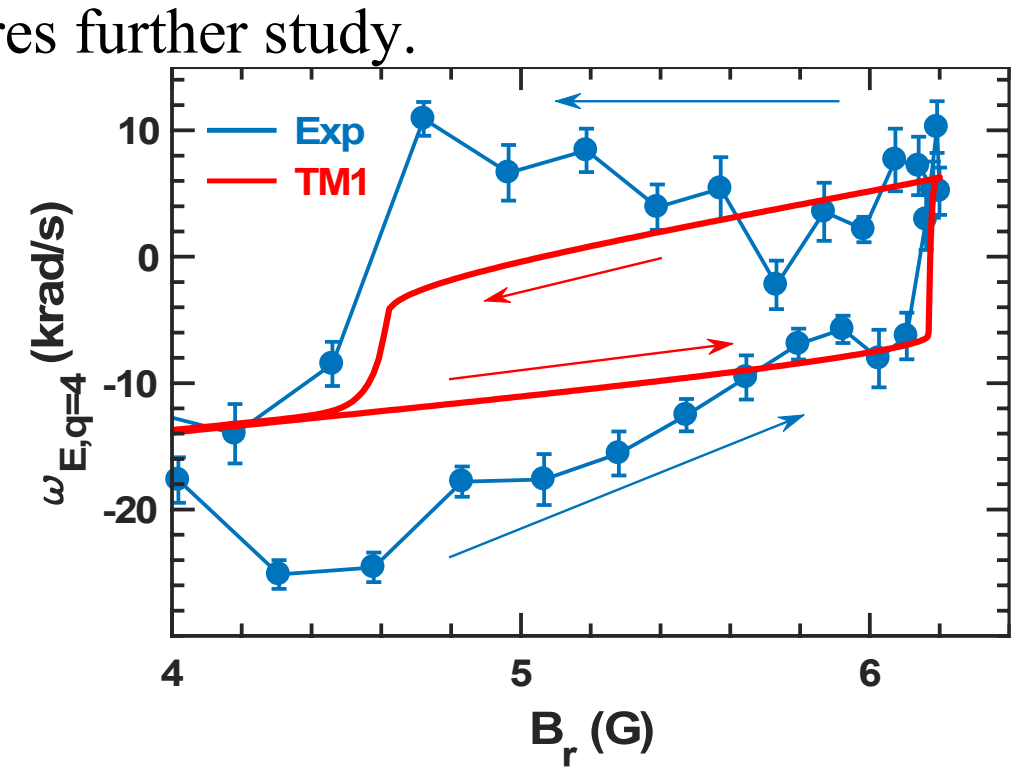

**Figure 20**. Comparison between experimental $\omega_E$ (blue) and TM1 simulation of $\omega_E$ at the $q$ = 4 surface (red curve) shown as a function of the $m/n$ =8/2 RMP amplitude $B_r$. Here, the evolution of $\omega_E$ in the first period of shot 158115 is shown.

### 3.6. Sensitivity to transport coefficients

According to the theoretical model of TM1, it is clear that both the penetration threshold and the resulting transport are sensitive to the transport coefficients at the rational surface. To confirm the sensitivity of the simulations to the choice of transport coefficients, simulations of 8/2 field penetration for the low density plasma #158115 is performed for a range of transport coefficients, $D_\perp = \mu = \chi_\perp$ = 0.5 m²/s (blue curve), 1 m²/s (red curve) and 2 m²/s (black curve), as shown in figure 21. It is found that smaller transport coefficients decrease the 8/2 penetration threshold and narrows the hysteresis loop, making field penetration easier to achieve, and vice versa for larger transport coefficients. The reason is that the ion polarization current caused by the RMP, affects the parallel transport of heat and particle as indicated by the first term in the right side of Equations (A4) and (A5) in Appendix A. The density and temperature will change more (less) for a smaller (larger) $D_\perp$ and $\chi_\perp$, leading to more (less) flattening in $n_e$ and $T_e$ and hence more (less) change in $\omega_{*e}$.

The effect of ion polarization current (parallel current perturbation) on electron continuity and electron energy transport is important in the dynamics of RMP penetration [13,69]. The polarization current term is turned off in one of our simulations to distinguish its effect, by setting $d_1 = 0$ (see Equation (A4) and (A5) in Appendix A), and the result is shown by the green dashed curve in figure 21. By turning off the polarization current term in Equations (A4) and (A5), $n_e$ and $T_e$ will not change significantly, resulting in a constant $\omega_{*e}$ and requiring a much larger RMP amplitude to force $\omega_{\perp e} = \omega_{*e} + \omega_E \approx 0$.

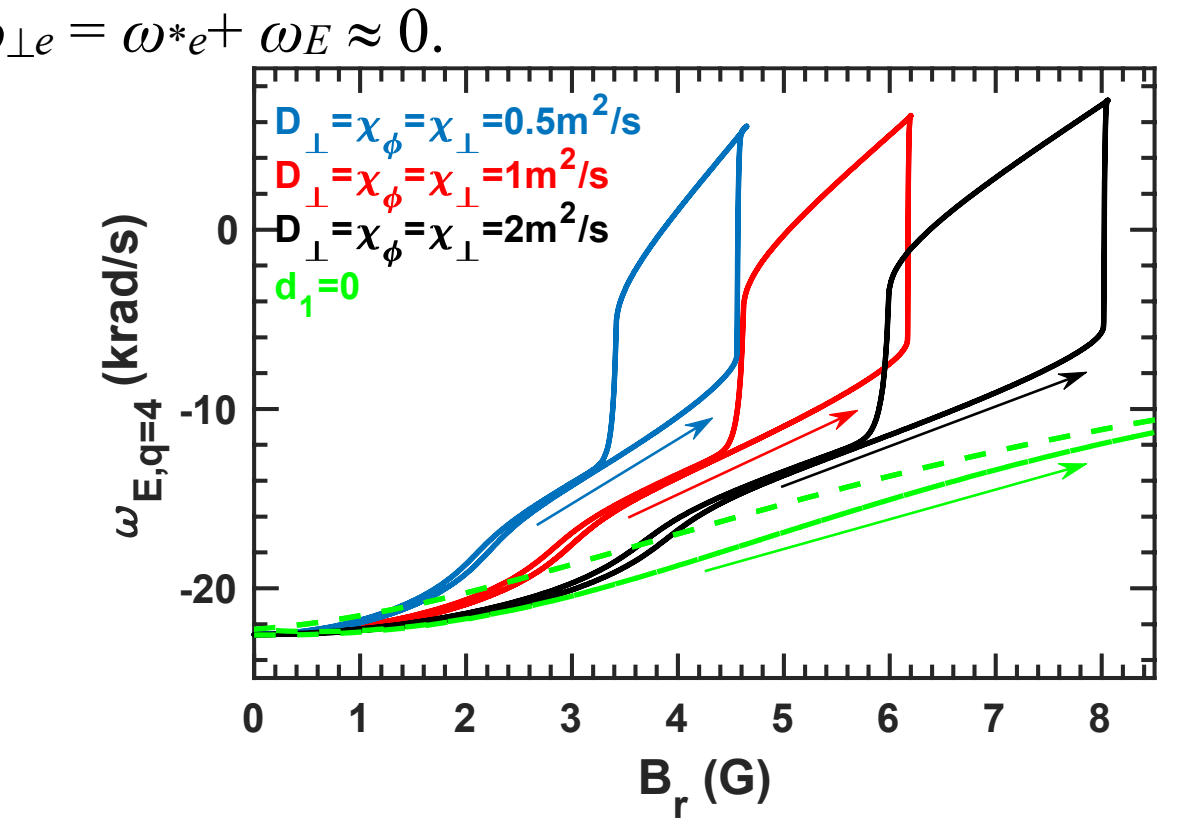

**Figure 21**. Modeled $\omega_E$ at $q = 4$ is shown as a function of RMPs amplitude $B_r$ for different transport coefficients, using the profiles in #158115.

Quantitatively, the variation in the penetration threshold is -25%/+50% for a -50%/+100% change in the transport coefficients, from figure 21. From our TRANSP analysis, the diffusivities show time slice to time slice variations of up to 20% due to variations in the data and profile fitting going into TRANSP. Therefore, we estimate that uncertainty in the penetration threshold to be ≈10%. This result is encouraging. However, to be more rigorous in our uncertainty quantification we would need to perform a forward modeling analysis using a statistical ensemble of initial conditions, which is currently beyond the scope of our capabilities. On the other hand, our analysis demonstrates the importance of using experimentally relevant parameters, including resistivity, collisionality, and transport coefficients, to obtain reliable results that can be compared profitably to experiment.

## 4. Discussion and Summary

The role of resonant field penetration on ELM suppression and density pump-out are investigated in this paper using nonlinear two-fluid MHD equations with a cylindrical circular tokamak geometry. This cylindrical geometry is different from the DIII-D highly shaped plasma. To address the effects of geometry, we use the GPEC code to calculate the ideal-MHD response of the plasma to the I-coils, with full shaping effects included. However, the toroidal mode coupling during the nonlinear stage is still ignored. Fortunately, the toroidal coupling between the 11/2 and 8/2 components from GPEC is quite small (<10%) and the saturated island width is also small (<2.5% of poloidal flux). These factors suggest that the ignored toroidal coupling in the penetration threshold and certainly on the saturated island width should be weak, however this issue awaits future study using nonlinear toroidal codes.

It is perhaps surprising that the TM1 model combined with GPEC captures the most ubiquitous attributes and trends in the plasma response to RMPs. The application of the TM1 model captures the most important features observed in DIII-D experiments concerning the phenomenology and scaling of ELM suppression and pump-out. What remains is to address the question of what additional physics is involved that will make an important contribution to predictive understanding at reactor scale. This latter question is not just philosophical as we may encounter important physics effects that arise at reactor scale even if they do not dominate at the level of the DIII-D experiments. We have pointed out that the ion thermal transport channel is not addressed by TM1 and we have seen that the impurity carbon-VI ion temperature at the top of the pedestal does not respond to island formation at the pedestal top. Also, the radial electric field response is not captured by TM1 outside of the $q = 4$ surface, and it is most likely that non-ambipolar and turbulent transport effects are involved in the change in $E_r$ before and after ELM suppression. Therefore, many details remain to be understood even though the most ubiquitous features of the DIII-D pump-out and ELM suppression data are well captured by TM1,

An important question is whether the TM1 model of density pump-out works much better than it should. What we mean here is that we calculate density flattening across a single magnetic island close to the separatrix, but it is more likely that a very narrow region ($\Delta\psi_N \approx 1\%$) around the separatrix is stochastic due to multiple edge harmonics near the separatrix. If the effect of stochasticity is essentially identical to that of an edge island for the parameters relevant to DIII-D and ITER then we may have an acceptable predictive capability for pump-out in ITER using TM1, but not necessarily a physically complete picture. How the physics of stochasticity close to the separatrix deviates from the simple TM1 pump-out predictions will be an important area of future study. For example, recent XGC simulations using linear M3D-C1 predictions demonstrate that pump-out arising from a combination of edge stochasticity near the separatrix and turbulence in the ETB [36]. These XGC simulations are not self-consistent and do not run to steady state, however the study points out some physical effects that may become more important at ITER scale.

For the island formation at the top of the pedestal, we expect that the physics contained in TM1 will hold for ITER for the penetration threshold and resulting electron pressure reduction at the pedestal top. It will be important in future studies to quantify the effect of the islands on ion thermal transport at the top of the pedestal using gyrokinetic simulation. There are indications from experiment [15] and gyrokinetic simulation [36] that enhanced turbulence could be playing a role, particularly in ion thermal transport [21]. Adding the magnetic island topology into these simulations will be an important future step.

In summary, we have addressed the key phenomenology of ELM suppression and density pump-out in low-collisionality DIII-D ISS plasmas using nonlinear two-fluid TM1 simulations. We find that:

1) The formation of magnetic islands at the foot of the pedestal produces experimentally relevant density pump-out. Pump-out weakens with increasing density due to the inverse dependence of the parallel particle transport on collisionality.
2) Resonant field penetration at the top of pedestal requires low-density in DIII-D ISS plasmas, providing a simple explanation for the ubiquitous observation of a low-density threshold for ELM suppression. Island formation at the top of the pedestal enhances particle and thermal transport, reducing the height and width of the electron pedestal sufficient for PBM stabilization and ELM suppression.
3) Strong temperature gradients persist in the ETB at low collisionality during ELM suppression and pump-out, due to the effective screening of resonant fields between the top and bottom of the pedestal at experimentally relevant resistivity. This screening is predicted to break down for resistivity 10X higher than in experiment, indicating that experimentally relevant levels of resistivity (and transport) must be used to make model comparisons with experimental data.
4) Higher RMP amplitude is required for field penetration at higher pedestal density and/or at lower co-$I_p$ rotation. A scaling is developed that predicts lower RMP amplitudes required for pedestal-top penetration in ITER due to the much lower diamagnetic frequency expected at the top of the ITER pedestal compared to present devices.

**Acknowledgements:** The authors would like to thank T.E. Evans, R. Groebner, N.M. Ferraro, Z. Lin, and R.A. Moyer for their helpful comments. Part of the data analysis was performed using the OMFIT integrated modeling framework [61]. This material is based upon work supported by the U.S. Department of Energy, Office of Science, Office of Fusion Energy Sciences, using the DIII-D National Fusion Facility, a DOE Office of Science user facility, under Awards DE-AC02-09CH11466, DE-FOA-0001386, DESC0015878 and DE-FC02-04ER54698.

the United States Government or any agency thereof. The views and opinions of authors expressed herein do not necessarily state or reflect those of the United States Government or any agency thereof.

## Appendix A. TM1 model

In the large aspect ratio approximation, the magnetic field is defined as $\boldsymbol{B} = B_{0t}\boldsymbol{e}_z + B_{0\theta}\boldsymbol{e}_\theta + \nabla\psi\times\boldsymbol{e}_z$, where $\psi$ is the flux function, and the perturbed helical flux $\psi_{m/n} = \Psi_{m/n}\exp[\mathrm{i}(m\theta + n\varphi)]$ . $B_{0t}$ (constant in radial direction) and $B_{0\theta}$ are the equilibrium magnetic field in the $\boldsymbol{e}_z$ (axial) and $\boldsymbol{e}_\theta$ (poloidal) direction, and the subscript 0 denotes an equilibrium quantity. The ion velocity $\boldsymbol{v} = v_\parallel\boldsymbol{e}_\parallel + \boldsymbol{v}_\perp$, where $v_\parallel$ and $\boldsymbol{v}_\perp = \nabla\phi\times\boldsymbol{e}_z$ are the parallel (to the magnetic field) and the perpendicular velocity, respectively. The cold ion assumption is made as in Ref. [48]. To obtain $\psi$, $v_\parallel$, $\boldsymbol{v}_\perp$, the electron density $n_e$ and temperature $T_e$, the electron continuity equation, the generalized Ohm's law, the equation of motion in the parallel and the perpendicular direction (after taking the operator $\boldsymbol{e}_z\cdot\nabla\times$), and the electron energy transport equation, are solved [47]. Normalizing the length to the minor radius $a$, the time $t$ to $\tau_R$, $\psi$ to $aB_{0t}$, $v$ to $a/\tau_R$ and $T_e$ and $n_e$ to their values at the magnetic axis, where $a$ is the minor radius and $\tau_R = a^2/\eta$ is the resistive time according to the Sauter [60] neoclassical resistivity $\eta$, these equations become [47]

$$\frac{d\psi}{dt} = E - \eta j + \Omega(\nabla_\parallel n_e + \nabla_\parallel T_e), \qquad \text{(A1)}$$

$$\frac{dv_\parallel}{dt} = -C_s^2\nabla_\parallel P/n_e + \mu\nabla_\perp^2 v_\parallel, \qquad \text{(A2)}$$

$$\frac{dU}{dt} = -S^2\nabla_\parallel j + \mu\nabla_\perp^2 U + S_m, \qquad \text{(A3)}$$

$$\frac{dn_e}{dt} = d_1\nabla_\parallel j - \nabla_\parallel(n_e v_\parallel) + \nabla\cdot(D_\perp\nabla n_e) + S_n, \qquad \text{(A4)}$$

$$\frac{3}{2}n_e\frac{dT_e}{dt} = d_1 T_e\nabla_\parallel j - T_e n_e\nabla_\parallel v_\parallel + n_e\nabla\cdot(\chi_\parallel\nabla_\parallel T_e) + n_e\nabla\cdot(\chi_\perp\nabla_\perp T_e) + S_p, \qquad \text{(A5)}$$

where $d/dt = \partial/\partial t + \boldsymbol{v}_\perp\cdot\nabla$, $U = -\nabla_\perp^2\phi$ is the plasma vorticity, $\mu$ the plasma viscosity, $\chi$ the heat conductivity and $D$ the particle diffusivity. $P = P_e = n_e T_e$ and the subscripts $\parallel$ and $\perp$ denote the parallel and the perpendicular components, respectively. $S_n$, $S_m$ and $S_p$ are the particle, momentum and heat source, which are derived from the equilibrium profiles based on Equations (A3), (A4) and (A5) and they are fixed in the time-dependent evolution. $E = \eta j_0$ is the equilibrium electric field derived from the original equilibrium current density $j_0$. The parameters in equations (A1)–(A5) are given by $d_1 = \omega_{ce}/\nu_{ei}$, $\Omega = \beta_e d_1$, $C_s = (T_e/m_i)^{1/2}/(a/\tau_R)$ and $S = \tau_R/\tau_A$, where $\beta_e = 4\pi n_e T_e/B_{0t}^2$, $\omega_{ce}$ is the electron cyclotron frequency, $\nu_{ei}$ is the electron-ion collisional frequency, and $\tau_A = a/V_A$ is the toroidal Alfven time.

The effect of the RMP is taken into account by the boundary condition

$$\psi_{m/n}|_{r=a} = \psi_a aB_{0t}\cos(m\theta + n\varphi), \qquad \text{(A6)}$$

where $\psi_a$ describes the normalized helical magnetic flux amplitude of the $m/n$ component at $r = a$. The radial magnetic field perturbation at $r = a$ is given by $b_{1r} = -m\psi_a B_{0t}\sin(m\theta + n\varphi)$. The full toroidal code GPEC [41] is used to evaluate the total of vacuum field and ideal plasma response $B_r$ at plasma edge for all relevant resonant components as above. Then the value of each component at plasma edge is obtained based on singular-value decomposition (SVD) and is used as the boundary condition for RMPs (similar to figure 4(c) in Ref. [79]). In the simulation, the island width $W$ is calculated based on the definition [78]

$$W = 4\left(\frac{|\psi_s|}{sr_s B_\theta(r_s)}\right)^{1/2} r_s, \qquad \text{(A7)}$$

where $\psi_s$, s, $r_s$ and $B_\theta$ are the helical flux perturbation, magnetic shear, radius of rational surface and equilibrium poloidal magnetic field at the rational surface of $q = m/n$. It is well known that the flux perturbation at the rational surface is non-zero (due to kink response) even though the applied RMP is shielded by the plasma [59,80].